\documentclass[twocolumn,resetfootnote,trackchanges]{aastex7}

\usepackage{textcomp,gensymb}
\usepackage{amsmath}
\usepackage{graphicx}
\graphicspath{{./}{figs/}}
\usepackage{subcaption}
\usepackage{tabularx}
\usepackage{makecell, booktabs}
\usepackage{enumitem}
\usepackage{tablefootnote}
\usepackage{threeparttable}
\usepackage{booktabs}  
\usepackage{bm}
\usepackage{xurl}
\usepackage{float}

\usepackage{CJK}
\DeclareTextSymbolDefault{\dh}{T1}

\newcommand\kms{$\textrm{km~s}^{-1}$}
\newcommand\ms{$\textrm{m~s}^{-1}$}

\newcommand\gcmcubed{$\textrm{g~cm}^{-3}$}

\newcommand\solmass{$M_{\odot}$}

\newcommand{\PSUAA}{Department of Astronomy \& Astrophysics, 525 Davey Laboratory, 251 Pollock Road, Penn State, University Park, PA, 16802, USA}
\newcommand{\PSUCEHW}{Center for Exoplanets and Habitable Worlds, 525 Davey Laboratory, 251 Pollock Road, Penn State, University Park, PA, 16802, USA}
\newcommand{\PSUARC}{Astrobiology Research Center, 525 Davey Laboratory, 251 Pollock Road, Penn State, University Park, PA, 16802, USA}
\newcommand{\PSETI}{Penn State Extraterrestrial Intelligence Center, 525 Davey Laboratory, 251 Pollock Road, Penn State, University Park, PA, 16802, USA}
\newcommand{\UA}{Steward Observatory, The University of Arizona, 933 N.\ Cherry Avenue, Tucson, AZ 85721, USA}

\newcommand{\Penn}{Department of Physics and Astronomy, University of Pennsylvania, 209 S 33rd St, Philadelphia, PA 19104, USA}
\newcommand{\Caltech}{Department of Astronomy, California Institute of Technology, 1200 E California Blvd, Pasadena, CA 91125, USA}

\newcommand{\GoddardESAL}{Exoplanets and Stellar Astrophysics Laboratory, NASA Goddard Space Flight Center, Greenbelt, MD 20771, USA}

\newcommand{\NOAO}{U.S. National Science Foundation National Optical-Infrared Astronomy Research Laboratory, 950 N.\ Cherry Ave., Tucson, AZ 85719, USA}

\newcommand{\Macquarie}{School of Mathematical and Physical Sciences, Macquarie University, Balaclava Road, North Ryde, NSW 2109, Australia}
\newcommand{\MacquarieCentre}{Astrophysics and Space Technologies Research Centre, Macquarie University, Balaclava Road, North Ryde, NSW 2109, Australia}

\newcommand{\JPL}{Jet Propulsion Laboratory, California Institute of Technology, 4800 Oak Grove Drive, Pasadena, California 91109}

\newcommand{\UCI}{Department of Physics \& Astronomy, The University of California, Irvine, Irvine, CA 92697, USA}
\newcommand{\Carleton}{Carleton College, One North College St., Northfield, MN 55057, USA}
\newcommand{\Carnegie}{Earth and Planets Laboratory, Carnegie Science, 5241 Broad Branch Road, NW, Washington, DC 20015, USA}
\newcommand{\PSUICS}{Institute for Computational and Data Sciences, Penn State, University Park, PA, 16802, USA}
\newcommand{\PSUCASt}{Center for Astrostatistics, 525 Davey Laboratory, 251 Pollock Road, Penn State, University Park, PA, 16802, USA}

\newcommand{\UAm}{Anton Pannekoek Institute for Astronomy, 904 Science Park, University of Amsterdam, Amsterdam, 1098 XH}

\newcommand{\Amherst}{Department of Physics and Astronomy, Amherst College, 25 East Drive, Amherst, MA 01002, USA}
\newcommand{\UWy}{Department of Physics \& Astronomy, University of Wyoming, Laramie, WY 82070, USA}

\begin{document}

\title{\textit{Searching for GEMS:} \\ The Occurrence of Giant Planets orbiting M-dwarfs within 100 pc}

\author[0000-0001-9816-0878]{Rowen I. Glusman}
\affiliation{\UAm}
\email{rowen.glusman@student.uva.nl}

\author[0000-0003-4835-0619]{Caleb I. Ca\~nas}
\altaffiliation{NASA Postdoctoral Fellow}
\affiliation{\GoddardESAL}
\email{c.canas@nasa.gov}

\author[0000-0001-8401-4300]{Shubham Kanodia}
\affiliation{\Carnegie}
\email{skanodia@carnegiescience.edu}

\author[0000-0002-7127-7643]{Te Han}
\affil{\UCI}
\email{teh2@uci.edu}

\author[0000-0002-3853-7327]{Rachel B. Fernandes}
\altaffiliation{President's Postdoctoral Fellow}
\affiliation{\PSUAA}
\affiliation{\PSUCEHW}
\email{rbf5378@psu.edu}

\author[0000-0001-7409-5688]{Gu\dh mundur Stef\'ansson}
\affil{\UAm} 
\email{g.k.stefansson@uva.nl}

\author[0000-0001-6545-639X]{Eric B. Ford}
\affiliation{\PSUAA}
\affiliation{\PSUCEHW}
\affiliation{\PSUCASt}
\affiliation{\PSUICS}
\email{ebf11@psu.edu}

\author[0000-0001-8222-9586]{Marissa Maney}
\affil{Harvard-Smithsonian Center for Astrophysics, 60 Garden Street, Cambridge, MA 02138, USA}
\email{marissa.maney@cfa.harvard.edu}

\author[0000-0002-0048-2586]{Andrew Monson}
\affil{\UA}
\email{andymonson@arizona.edu}

\author[0009-0000-1825-4306]{Andrew Hotnisky}
\affiliation{\PSUAA}
\affiliation{\PSUCEHW}
\email{amh7996@psu.edu}

\author[0000-0001-9596-7983]{Suvrath Mahadevan}
\affiliation{\PSUAA}
\affiliation{\PSUCEHW}
\affiliation{\PSUARC}   
\email{suvrath@astro.psu.edu}

\author[0009-0009-4977-1010]{Michael Rodruck}
\affil{Department of Physics, Engineering, and Astrophysics, Randolph-Macon College, Ashland, VA 23005, USA}
\email{mrodruck@gmail.com} 

\author[0000-0001-5847-9147]{Kristo Ment}
\affiliation{\PSUAA}
\affiliation{\PSUCEHW}
\email{kxm821@psu.edu}

\author[]{Andrew McWilliam}
\affiliation{The Observatories of the Carnegie Institution for Science, 813 Santa Barbara Street, Pasadena, CA 91101, USA}
\email{andy@carnegiescience.edu}

\author[0000-0001-9662-3496]{William D. Cochran}
\affil{McDonald Observatory and Department of Astronomy, The University of Texas at Austin}
\affil{Center for Planetary Systems Habitability, The University of Texas at Austin}
\email{wdc@astro.as.utexas.edu}

\author[0000-0001-8020-7121]{Knicole D. Col\'on}
\affil{\GoddardESAL}
\email{knicole.colon@nasa.gov}

\author[0000-0002-0078-5288]{Mark R.~Giovinazzi}
\affiliation{\Amherst}
\email{mgiovinazzi@amherst.edu}

\author[0000-0003-0353-9741]{Jaime A. Alvarado-Montes}
\affiliation{Australian Astronomical Optics, Macquarie University, Balaclava Road, North Ryde, NSW 2109, Australia }
\affiliation{\MacquarieCentre}
\email{jaime.alvaradomontes@mq.edu.au}

\author[0000-0003-4384-7220]{Chad F.\ Bender}
\affiliation{\UA}
\email{cbender@arizona.edu}

\author[0000-0002-6096-1749]{Cullen H.\ Blake}
\affiliation{\Penn}
\email{chblake@sas.upenn.edu}

\author[0009-0006-3467-630X]{Alexandra Boone}
\affil{\UWy}
\email[]{aboone1@uwyo.edu}

\author[0000-0002-2144-0764]{Scott A. Diddams}
\affil{Electrical, Computer \& Energy Engineering, University of Colorado, 1111 Engineering Dr.,  Boulder, CO 80309, USA}
\affil{Department of Physics, University of Colorado, 2000 Colorado Avenue, Boulder, CO 80309, USA}
\email[]{scott.diddams@colorado.edu}  

\author[0000-0002-5463-9980]{Arvind F.\ Gupta}
\affil{\NOAO}
\email{arvind.gupta@noirlab.edu}

\author[0000-0003-1312-9391]{Samuel Halverson}
\affiliation{\JPL}
\email{samuel.halverson@jpl.nasa.gov}

\author[0000-0001-9626-0613]{Daniel Krolikowski}
\affil{\UA}
\email{krolikowski@arizona.edu}

\author[0000-0002-9082-6337]{Andrea S.J. Lin}
\affil{\Caltech}
\email{asjlin@caltech.edu}

\author[0000-0001-8720-5612]{Joe P.\ Ninan}
\affil{Department of Astronomy and Astrophysics, Tata Institute of Fundamental Research, Homi Bhabha Road, Colaba, Mumbai 400005, India}
\email{joe.ninan@tifr.res.in}

\author[0000-0003-0149-9678]{Paul Robertson}
\affil{\UCI}
\email{paul.robertson@uci.edu}

\author[0000-0001-8127-5775]{Arpita Roy}
\affiliation{Astrophysics \& Space Institute, Schmidt Sciences, New York, NY 10011, USA}
\email{arpita308@gmail.com}

\author[0000-0002-4046-987X]{Christian Schwab}
\affil{\Macquarie}
\email{mail.chris.schwab@gmail.com}

\author[0000-0002-4788-8858]{Ryan Terrien}
\affiliation{\Carleton}
\email{rterrien@carleton.edu}

\author[0009-0008-2801-5040]{Johanna Teske}
\affiliation{\Carnegie}
\email{jteske@carnegiescience.edu}

\author[0000-0001-6160-5888]{Jason T.\ Wright}
\affiliation{\PSUAA}
\affiliation{\PSUCEHW}
\affiliation{\PSETI}
\email{astrowright@gmail.com}

\correspondingauthor{Rowen I. Glusman}
\email{rowen.glusman@student.uva.nl}

\begin{abstract}

We present results from a systematic search for transiting short-period Giant Exoplanets around M-dwarf Stars (GEMS; $P < 10$ days, $R_p \gtrsim 8~R_\oplus$) within a distance-limited 100 pc sample of $149,316$ M-dwarfs using TESS-Gaia Light Curve (TGLC) data. We describe the development and application of the \texttt{TESS-miner} package and associated vetting procedures used in this analysis. To assess detection completeness, we conducted $\sim$72 million injection-recovery tests across $\sim$26,000 stars with an average of $\sim$3 sectors of data per star, subdivided into early-type (M0--M2.5), mid-type (M2.5--M4), and late-type (M4 or later) M-dwarfs. Our pipeline demonstrates high sensitivity across all subtypes within the injection bounds.

We estimate the occurrence rates of short-period GEMS as a function of stellar mass, and combine our measured rates with those derived for FGK stars, fitting an exponential trend with stellar mass, consistent with core-accretion theory predictions. We find GEMS occurrence rates of $0.118\% \pm 0.068\%$ for early-type M-dwarfs, $0.153\% \pm 0.069\%$ for mid-type, and $0.036\% \pm 0.024\%$ for late-type M-dwarfs, with a mean rate of $0.068\%\pm0.024\%$ across the full sample. While our search spanned $1.0~\mathrm{days} < P < 10.0$ days, these rates were calculated using planets orbiting with $1.0~\mathrm{days} < P < 5.0$ days. This work establishes the basis for future occurrence rate studies of transiting GEMS.

\end{abstract}

\keywords{extrasolar gaseous giant planets (509) — hot Jupiters (753) — M-dwarf stars (982) —  planet hosting stars (1242) — binary stars (154) — companion stars (291) — eclipsing binary stars (444) — algorithms (1883) — radial velocity (1332) — transit photometry (1709)}

\section{Introduction} \label{sec:intro}
M-dwarfs are the most abundant stellar type in the Milky Way \citep{reid_low-mass_1997, henry_solar_2006, reyle_10_2021}, with masses ranging from $0.08~\mathrm{M_\odot} \lesssim M_\star \lesssim 0.6~\mathrm{M_\odot}$ and effective temperatures between $2600$–$4000~\mathrm{K}$ \citep[][]{pecaut_intrinsic_2013}. Their protoplanetary disks are correspondingly less massive than those of higher-mass stars, as shown by disk mass distributions in previous studies \citep[e.g.,][]{andrews_mass_2013, pascucci_steeper_2016, Manara23}. Due to their lower masses, M-dwarfs also have longer Keplerian orbital timescales at a given orbital distance than more massive stars.

M-dwarfs frequently host multiple terrestrial planets \citep{Mulders2015, dressing_occurrence_2015, gaidos_they_2016, hardegree_ullman_kepler_2019, hsu_occurrence_2020}, but the occurrence rate and dominant formation pathway of \uline{G}iant \uline{E}xoplanets around \uline{M}-dwarf \uline{S}tars (GEMS) remain uncertain \citep[e.g.,][]{endl_exploring_2006, johnson_giant_2010, maldonado_connecting_2019, schlecker_rv-detected_2022, gan_occurrence_2023, bryant_occurrence_2023}. According to core-accretion theory, giant planets form during the protoplanetary phase through i.) grain condensation, ii.) planetesimal coagulation \citep{Greenberg1978_collisional_evol, Wetherill1989, Aarseth1993, Kokubo1996, Kokubo02}, iii.) oligarchic growth of planetary embryos \citep{Kokubo1998, Kokubo2000}, and iv.) runaway gas accretion onto solid cores \citep{Mizuno1980, bodenheimer_calculations_1986, Pollack96, ikoma_formation_2000}. However, the low disk masses and long Keplerian timescales around M-dwarfs are expected to inhibit this process \citep{laughlin_core_2004}. Alternatively, GEMS may form via gravitational instability within the protoplanetary disk, whereby regions of the disk implode directly and form gas giants without solid cores \citep{boss_giant_1997, boss_forming_2023}. Regardless of the formation mechanism, GEMS must originate at large orbital separations --- due to the implausibility of in-situ formation at their current locations --- and subsequently migrate inward to avoid engulfment by the host star \citep{boss_formation_2006, boss_formation_2011}. Atmospheric characterization has been proposed as a method to distinguish between formation pathways, but more GEMS discoveries are needed to enable such studies \citep{helled_metallicity_2010, oberg_effects_2011, madhusudhan_toward_2014, knierim_constraining_22}. 


Although transiting GEMS discovered to date have primarily been detected around early M-dwarfs (M0-M2), six GEMS transiting mid M-dwarfs have recently been confirmed: TOI-5205 b \citep{kanodia_toi-5205b_2023}, TOI-3235 b \citep{hobson_toi-3235_2023}, TOI-519~b \citep{parviainen_toi-519_2021, kagetani_mass_2023, hartman_toi_2023}, TOI-4860 b \citep{almenara_toi-4860_2023, triaud_m_2023}, TOI-6894 b \citep{Bryant2025}, and TOI-7149~b \citep{kanodia_toi-7149_2025}. Radial velocity (RV) surveys have confirmed the presence of non-transiting GEMS orbiting mid to late M-dwarfs as well \citep[e.g.,][]{morales_giant_2019, quirrenbach_carmenes_2022-1}.\footnote{We note here that non-transiting GEMS have a $M_p\sin{i}$ degeneracy in that their exact orbital inclination cannot be derived from RV observations alone. Therefore, they may still prove to be substellar objects.} Furthermore, the Gaia mission has begun to detect candidate GEMS on long-period orbits using the astrometric technique \citep{holl_gaia_2023, gaia_collab_2023_stellar_mult} that can be confirmed using radial velocities \citep[e.g.,][]{Stefansson2025}. The presence of GEMS orbiting these
low-mass stars exceeds the inferred mass budget of solids in protoplanetary disk samples, challenging the predictions of population synthesis models \citep{burn_new_2021, kanodia_toi-5205b_2023}.

Before the launch of the Transiting Exoplanet Survey Satellite \citep[TESS;][]{Ricker14}, the Kepler mission \citep{Borucki10} was designed to detect the transits of Earth-sized planets orbiting Sun-like stars, and was only able to achieve the necessary S/N for targets down to $V$-mag $\sim14$. As a result, the primary target sample included relatively few M-dwarfs, which are generally fainter than this magnitude in the Kepler field of view. Although GEMS-specific studies using ground-based data were able to place upper limits on their occurrence rates \citep{kovacs_hot_2013, Zendejas2013, Obermeier2016}, the discoveries from both Kepler and ground-based surveys were too limited to produce reliable constraints.

Most GEMS detections to date are from RV surveys of nearby M-dwarfs \citep{endl_exploring_2006, johnson_giant_2010,bonfils_harps_2013, maldonado_connecting_2019, sabotta_carmenes_2021, schlecker_rv-detected_2022, pinamonti_hades_2022,mignon2025}, which have largely confirmed the rarity of these planets by placing upper limits on their occurrence rates. Since 2018, TESS \citep{Ricker14} has observed millions of M-dwarfs, uncovering numerous transiting candidate GEMS \citep{guerrero_tess_2021}. Many of these are excellent targets for follow-up observations aimed at confirming their planetary nature and ruling out false positives such as eclipsing binaries (EBs) and brown dwarfs (BDs).

To discover GEMS and measure their occurrence rate, \citet{kanodia2024} introduced the motivation and framework for the \textit{Searching for GEMS} survey using TESS data, including the description of a 200 pc M-dwarf sample. Their goal was to enable occurrence rate measurements as a function of stellar mass for a distance-limited sample, which could then be compared to trends in protoplanetary disc mass \citep{pascucci_steeper_2016, Manara23} and planetary bulk properties \citep{Muller_2024, Kanodia_2025}. The present study constitutes the first analysis within that framework, focusing on the 100 pc subset of the sample defined by \cite{kanodia2024}. The exact selection criteria and sample boundaries are described in Section \ref{sec:sample}.

Previous occurrence rate studies of GEMS \citep{gan_occurrence_2023, bryant_occurrence_2023} have focused on samples which were biased toward earlier-type M-dwarfs due to limiting cuts on TESS sector, stellar mass, and/or stellar effective temperature (see Section \ref{sec:sample}, \autoref{tab:comparison}). If core-accretion theory does govern the dominant formation mechanism for GEMS, such a bias would be expected to yield higher occurrence rates (we address this sampling bias in Section \ref{sec:sample}). These studies were also limited by a high incidence of false positive (FP) candidates, often arising from EBs. High FP rates also contribute to the overestimation of the occurrence rates. In this work, we derive occurrence rates for GEMS across the full M-dwarf mass range and perform astrophysical false positive validation via RV follow-up for all of our candidates (see Section \ref{sec:specvetting}) to remove false positives and ensure a robust occurrence rate analysis.

In Section \ref{sec:sample}, we describe the sample selection for our 100 pc M-dwarf sample. In Section \ref{sec:pipeline_overview}, we explain the steps of our transit detection package \texttt{TESS-miner}, including data preparation, light curve detrending, transit fitting, and quality checks. Section \ref{sec:vetting} follows with a description of our automated, manual, and spectroscopic vetting techniques and presents a vetted and validated list of all GEMS detected within our sample. In Section In Section \ref{sec:overall_occ}, we compute the detection efficiency and completeness of our survey in planetary radius and orbital period space for early-, mid-, and late-type hosts using injection and recovery tests, where synthetic transit signals are added to real light curves and analyzed identically to real data to evaluate the sensitivity of the pipeline. We also motivate our methods for calculating the occurrence rates of GEMS across early-, mid-, and late-type M-dwarfs and present our results. In Section \ref{sec:discussion}, we place these results in the context of previous studies. Finally, we summarize our findings in Section \ref{sec:summary}.

\section{Sample Characteristics} \label{sec:sample}

\begin{figure*}
    \centering
    \includegraphics[width=\textwidth]{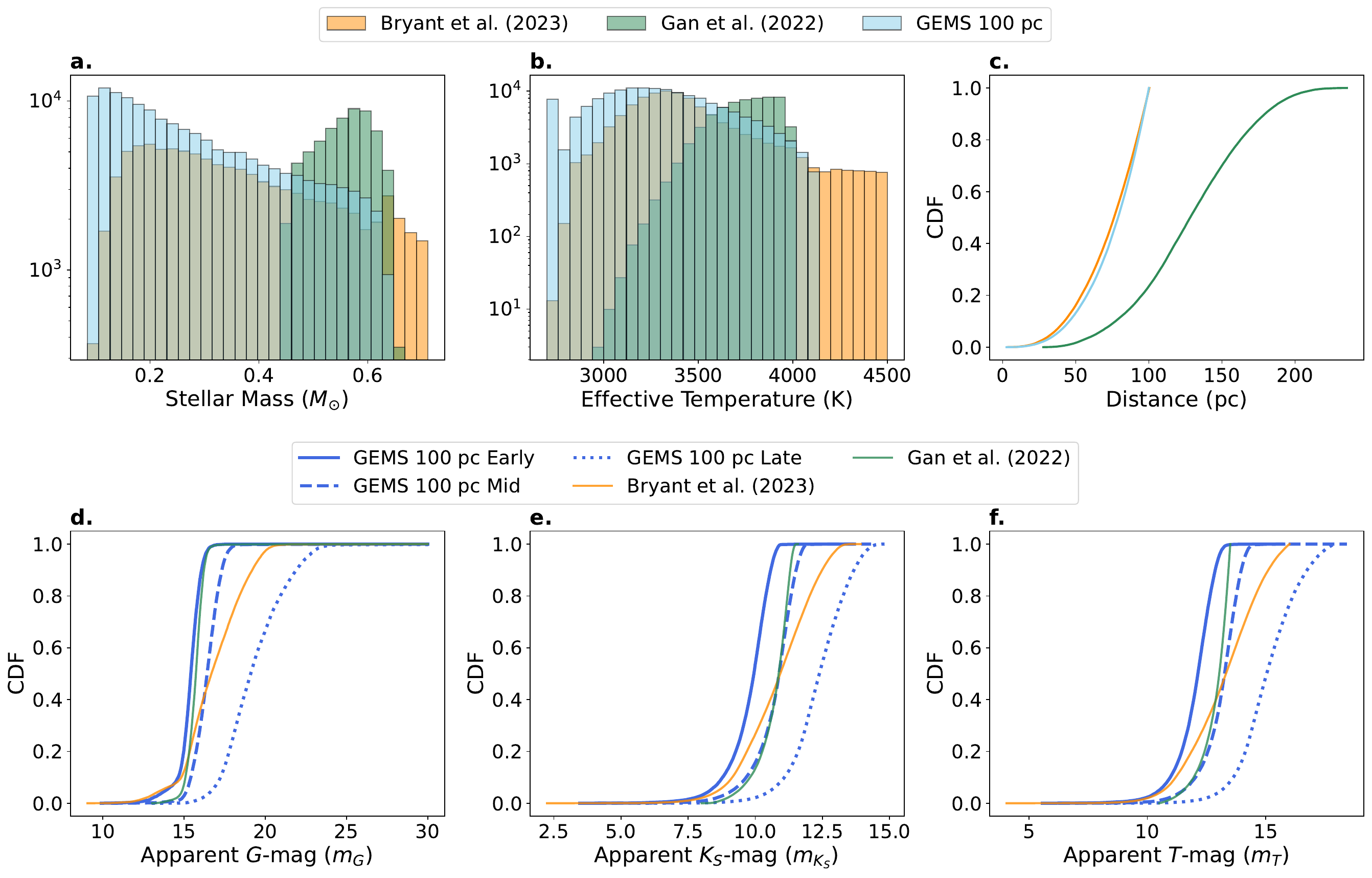}
    \caption{The stellar characteristics of the GEMS 100 pc sample compared with those of \citet{bryant_occurrence_2023} in orange and \citet{gan_occurrence_2023} in green. They are as follows: (a) stellar mass, (b) effective temperature, and the cumulative distribution functions (CDFs) of (c) distance, (d) $G$ mag, (e) $K_S$ mag, and (f) $T$ mag. In the latter three, the GEMS 100 pc sample (in dark blue) is divided into early (solid), mid (dashed), and late (dotted).}
    \label{fig:surveyfig}
\end{figure*}

\begin{deluxetable*}{|p{2.5cm}|p{4.0cm}|p{5.0cm}|p{4.5cm}|}
\tablecaption{A comparison of (a) the cuts used to define, and (b) the descriptive metrics of \citet{bryant_occurrence_2023}, \citet{gan_occurrence_2023}, and this work.\label{tab:comparison}}
\tablehead{
\hline
\multicolumn{4}{|c|}{\textbf{Cuts Applied to the Sample}} \\
\hline
\rule{0pt}{2.8ex} & \citet{bryant_occurrence_2023} & \citet{gan_occurrence_2023} & This work
}
\startdata
\rule{0pt}{2.8ex}Cuts & TESS Sectors 1--26 & TESS Sectors 1--26 & TESS Sectors 1--55 \\
\rule{0pt}{2.8ex} & $T_{\text{eff}} \leq 4500$ K & $2900 \leq T_{\text{eff}} \leq 4000$ K & $4.7 < M_{K_S} < 10.1$ \\
\rule{0pt}{2.8ex} & $R_\star/R_\odot \leq 0.75$ & $0.45 \leq M_\star/M_\odot \leq 0.65$ & $0.8 < J-K < 1.25$ \\
\rule{0pt}{2.8ex} & $d \leq 100$ pc & $T_{\text{mag}} \leq 13.5$ & $\varpi > 10$ mas \\
\rule{0pt}{2.8ex} & $T_{\text{mag}} \leq 16$ & $T_{\text{mag}} \leq 10.5 \cap D_{\text{TESS}} > 0.3$ & $\sigma_\varpi < 1$ mas \\
\rule{0pt}{2.8ex} & -- & $4.5 < M_{K_S} < 10.0$ & -- \\
\hline
\multicolumn{4}{|c|}{\textbf{Descriptive Metrics}} \\
\hline
\rule{0pt}{2.8ex}$N_{\text{stars}}$ & 91,306 & 60,819 & 149,316 \\
\rule{0pt}{2.8ex}Light Curves & TESS-SPOC\textsuperscript{a} & QLP\textsuperscript{b} & TGLC\textsuperscript{c} \\
\rule{0pt}{2.8ex}Distance & $0 \leq d \leq 100$ pc & $0 \leq d \leq 250$ pc & $0 \leq d \leq 100$ pc \\
\hline
\enddata
\tablenotetext{a}{\url{https://archive.stsci.edu/hlsp/tess-spoc} (TESS-SPOC; \citealt{caldwell_tess_2020})}
\tablenotetext{b}{\url{https://archive.stsci.edu/hlsp/qlp} (QLP; \citealt{huang_photometry_2020})}
\tablenotetext{c}{\url{https://archive.stsci.edu/hlsp/tglc} (TGLC; \citealt{Han2023})}
\end{deluxetable*}

We draw our M-dwarf sample from the 200 pc distance-limited catalog described in \citet{kanodia2024}, and observed in the TESS primary mission (PM; TESS sectors 1 -- 26) and first extended mission (EM1; TESS sectors 27 -- 55). The 100 pc subset is designated as a proving ground for the methods in this work, with the full 200 pc sample reserved for a future search. We queried our targets from Gaia DR3 \citep{gaia_collaboration_gaia_2023}, according to the same parallax error, quality, color, and $M_{K_s}$ criteria\footnote{We did not impose a cut on Gaia’s Renormalized Unit Weight Error (RUWE) or any other astrometric quality metric, as doing so could have biased our faint sample toward brighter, better-fitted targets.} as laid out in Section 4.3.1 of \citet{kanodia2024}. Following \citet{Moe2021}, we avoid a simple magnitude-limited search, which can bias a sample toward close, bright binaries. Such objects are already removed from RV target lists, artificially boosting hot-Jupiter occurrence rates relative to transit-based surveys such as this one. For the 100 pc sample, we include M-dwarfs with:

\begin{enumerate}
    \item Parallax $\varpi > 10$ mas,
    \item Parallax error $\sigma_\varpi < 1$ mas,
    \item Absolute $K$ magnitude $4.7 < M_{K_s} < 10.1$
    \item Color $0.8 < J-K < 1.25$, and
    \item Null flags indicating that the target is not in the \textit{Gaia} DR3 \texttt{qso\_candidates}, \texttt{galaxy\_candidates}, or \texttt{nss\_two\_body\_orbit} tables \citep{gaia_collab_2023_stellar_mult}.
\end{enumerate}

These cuts resulted in a total of 149,316 unique stars. We used the absolute $M_{K_S}$ bounds from \citet{kanodia2024} to split this sample into $27,995$ early ($<$ M2.5; $4.7 < M_{K_S} < 6.0$), $37,948$ mid (M2.5 -- M4; $6.0 < M_{K_S} < 7.1$), and $79,669$ late M-dwarfs ($>$ M4; $7.1 < M_{K_S} < 10.1$). The masses for all stars in our sample were calculated using the $M_K$ relations from Equation 4 in \citet{Mann19}, and the radii and effective temperatures ($T_{\text{eff}}$) using Equation 4 in \citet{mann_how_2015} (see \citet{mann_how_2015} Table 1 for the coefficients associated with the calculation of $R_*$ and $T_\text{eff}$).\footnote{Our sample slightly exceeds the calibrated range of the $M_{K_S}$–$T_\text{eff}$ relation. For stars with $9.1 < M_{K_S} < 10.1$, we assign $T_\text{eff} = 2700~\mathrm{K}$ by convention, which produces the spike seen in \autoref{fig:surveyfig}b.} We do not correct for reddening when creating the sample. While extinction is generally minimal within 100 pc \citep{Reis2011, Gontcharov2017}, we note that some regions may still be affected by modest reddening, which could introduce small biases in $M_K$-based stellar parameters. A table of our 100 pc sample, along with a table of all targets surviving until manual vetting (see Section 4.2), is available online\footnote{ \url{https://zenodo.org/records/15738847}}.
 
\autoref{fig:surveyfig} compares our sample to those from \citet{gan_occurrence_2023} and \citet{bryant_occurrence_2023}, two previous TESS-based searches for transiting GEMS. \autoref{tab:comparison} summarizes key differences in sample selection, cadence, and sector coverage. Our sample most closely resembles that of \citet{bryant_occurrence_2023}.  \citet{gan_occurrence_2023} notably extended their search to 200 pc, beyond the 100 pc limit adopted here and by \citet{bryant_occurrence_2023}. As shown in \autoref{fig:surveyfig}a and b, our sample includes a significantly more complete set of low-mass stars ($M_\star / M_\odot < 0.4$) than either prior study. This is largely due to the fact that the MIT Quick Look Pipeline used by \citet{gan_occurrence_2023} (QLP; \citet{huang_photometry_2020}) and the TESS Science Processing Operations Center Pipeline used by \citet{bryant_occurrence_2023} (TESS-SPOC; \citet{caldwell_tess_2020}) have more restrictive magnitude limits than the TESS-Gaia Light Curve \citep[TGLC;][]{Han2023} pipeline which we use (see Section \ref{sec:data_prep}.) 

\begin{figure*}
    \centering
    \includegraphics[width=\textwidth]{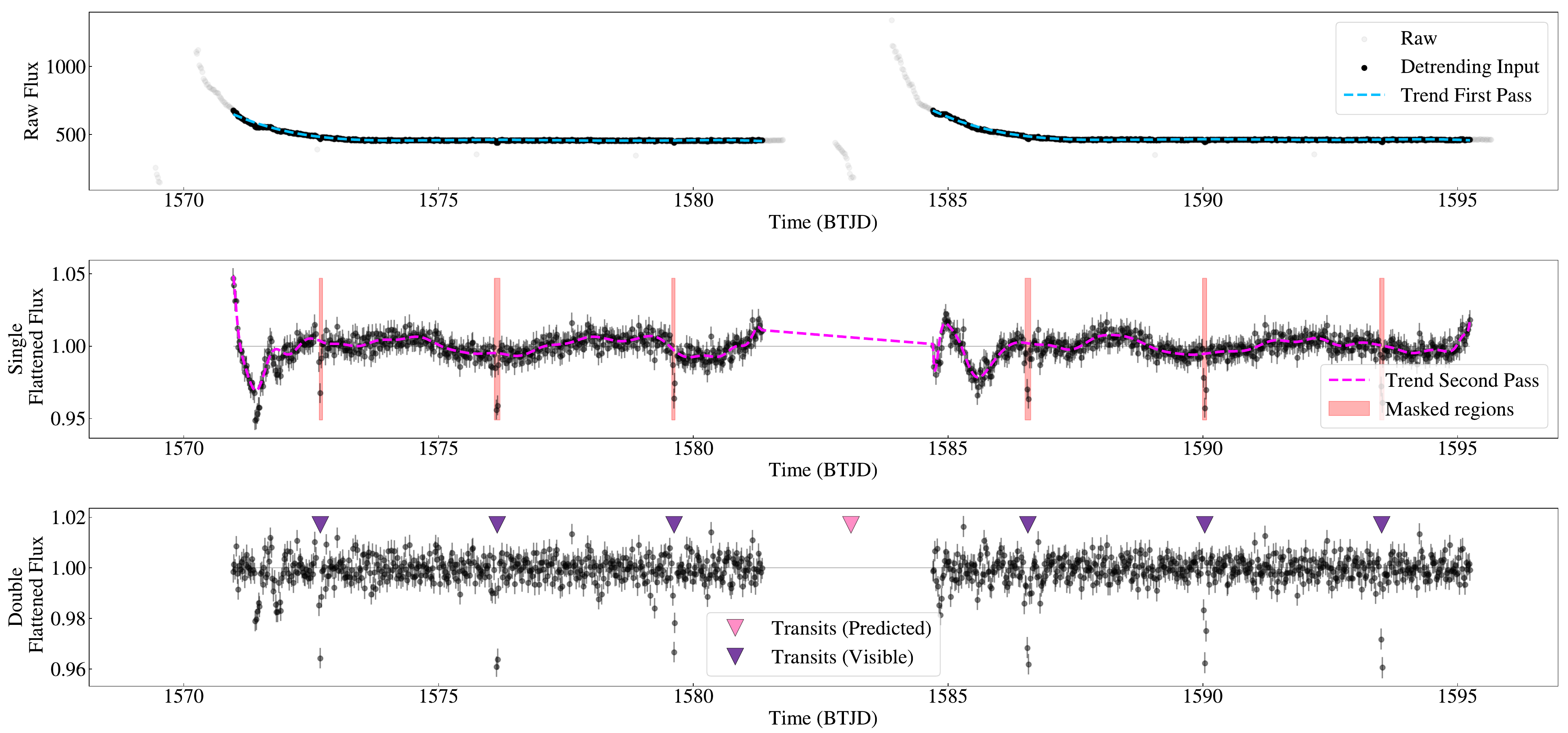}
    \caption{\textbf{Top:} The raw \texttt{aperture\_flux} calculated by TGLC for TIC 178709444 (TOI-762A) from Sector 10 plotted with the \texttt{wotan} cosine trend (dashed cyan) of window length 5.0 days. Light-gray points are regions masked (e.g., due to significant background contamination or non-zero data quality flags).  \textbf{Middle:} The single-detrended TGLC light curve with transits masked (light red) prior to the calculation of the second cosine trend (dashed magenta) with window length 0.5 days. \textbf{Bottom:} The double-detrended TGLC light curve marking transits detected by the BLS algorithm with purple triangles, and predicted transits (not contained in the data) indicated by pink triangles.}
    \label{fig:detrending_example}
\end{figure*}

We applied the same selection criteria that define our target sample to the NASA Exoplanet Archive\footnote{\url{https://exoplanetarchive.ipac.caltech.edu/}} \citep{Christiansen2025} to retrieve all confirmed GEMS, adopting $0.7 < R_p/R_J < 1.5$ and $1.0 < P/\mathrm{day} < 10.0$ as our working definition.\footnote{The lower radius limit corresponds to $R_p-\sigma(R_p) \ge 8.0~R_\oplus$; we quote radii in $R_J$ to one decimal place hereafter.}
As of this writing, the Archive lists 20 confirmed GEMS that (i) lie within 200 pc, (ii) satisfy our magnitude cuts, and (iii) were observed in PM and/or EM1. We also included TOI-6894 b ($d \simeq 73$ pc), recently confirmed by \citet{Bryant2025}. We selected the thresholds within our pipeline (Section \ref{sec:pipeline_overview}, Section \ref{sec:vetting}) so that all of the confirmed GEMS advanced to the manual vetting stage.

Of the 20 confirmed GEMS, seven lie within 100 pc. We retained these objects in our target list to assess the recovery performance of our pipeline on known systems.

\section{Transit Photometry with \texttt{TESS-miner}} \label{sec:pipeline_overview}

\subsection{Preparation of \texttt{TGLC} Light Curves} \label{sec:data_prep}

We use TGLCs \citep{Han2023} for our analysis in place of light curves from the QLP \citep{huang_photometry_2020} or TESS-SPOC \citep{caldwell_tess_2020}. TESS-SPOC targets are selected with apparent $T$-mag $m_\text{T}<12$, with some extension to 13 depending on mission priorities \citep{sullivan15, stassun18}. Meanwhile QLP targets are selected to have $m_\text{T}<16-17$ \citep{huang_photometry_2020}. TGLC light curves exist out to $m_\text{T} \approx 18$, allowing our sample fuller coverage across faint M-dwarfs. Moreover, while the \texttt{eleanor} package \citep{feinstein_eleanor_2019} provides access to light curves for stars not covered by QLP or TESS-SPOC, it does not correct for dilution in the photometry due to nearby stars. This omission leads to significantly underestimated transit depths and biases in planet radius estimates for crowded fields. Previous work has documented these issues: for example, \citet{kanodia_toi-5205b_2023} and \citet{kanodia_searching_2024-1} demonstrate that \texttt{eleanor}-based fits can mischaracterize transits in the presence of nearby stars. Meanwhile the TGLC pipeline, styled \texttt{tglc}, deblends crowding effects and provides complete sector-by-sector coverage. It has also demonstrated statistically accurate transit depths after correcting dilutions, making it well suited for robust occurrence rate studies across the TESS field \citep{Han2025_larger}. 

We retrieved 477,999 TGLCs from the TESS Primary Mission (PM) and First Extended Mission (EM1), covering all 149,316 stars in our sample.\footnote{The processing of the TESS second extended mission (EM2) with \texttt{tglc} is ongoing. At the time of writing, the light curves analyzed in this study through Sector 54 are available on MAST at \url{https://archive.stsci.edu/hlsp/tglc}.} We generated these light curves using the \texttt{tglc} package, applying the same processing parameters as the publicly available TGLC products hosted by the Mikulski Archive for Space Telescopes High-Level Science Products \citep{tglc_hlsp}. Each light curve was extracted from a $150 \times 150$ pixel full-frame image (FFI) cutout. The substantial $150 \times 150$ pixel size provides a robust estimate of the TESS point spread function (PSF). At each epoch, \texttt{tglc} fitted a PSF model based on stellar positions and magnitudes from Gaia Data Release 3 \citep{gaia_collaboration_gaia_2023}. \texttt{tglc} then decontaminated the FFI cutouts by subtracting a simulated image constructed solely from field stars and background flux located within the FFI cutout. For our analysis, we opted not to use the calibrated aperture flux (\texttt{cal\_aper\_flux}) in order to perform our own independent detrending steps. We instead used the aperture flux (\texttt{aperture\_flux}), computed by summing and normalizing the flux of the central $3 \times 3$ pixels of the decontaminated FFI cut at each timestamp. 

We analyzed each TESS sector independently to avoid mixing datasets with differing noise properties or cadences. Within each sector, we masked all timestamps where the target flux dropped below 25\% of the background, as such conditions preclude reliable signal detection. If fewer than 25\% of the original data points remained after masking, we flagged the sector as “Bad Data” and excluded it from analysis. As shown in Figure~\ref{fig:surveymap}, most of these excluded light curves lie near the crowded Galactic bulge, where contamination due to scattered light from the Earth and moon is common.

Our masking procedure, combined with regular TESS downlink gaps, created gaps in the time series. We defined a major gap as any interruption longer than 12 hours and masked intervening segments shorter than 24 hours between two such gaps. Additionally, we trimmed the first and last 0.4 days (9.6 hours) of each light curve to remove systematics associated with Earth rising above the satellite’s sunshade. While this effect primarily impacts Northern Hemisphere sectors, we applied the same cut to Southern data to suppress potential artifacts from thermal settling, pointing jitter, or scattered light, and to ensure consistent treatment across the sample.

Next, we took the following steps:

\begin{enumerate}
    \item \textbf{Initial Masking and Smoothing:} We began by generating a smoothed flux curve using a Savitzky-Golay filter \citep{savitzky1964smoothing} with a window length of 13 cadences and a cubic polynomial.
    \item \textbf{Residual and Sigma Calculation:} We then computed the residuals between the observed flux values and the smoothed flux. We also computed the standard deviation ($\sigma$) of the light curve.
    \item \textbf{Asymmetric Sigma Clipping:} We classified data points as outliers based on asymmetric sigma thresholds. Specifically, we masked out points if their residuals were either more than 3-$\sigma$ higher, or more than 10-$\sigma$ lower. This asymmetry ensures that strong positive deviations (e.g., flares) are removed more aggressively, preserving negative deviations such as planet transits.
    \item \textbf{Iterative Refinement:} We repeated the process for up to 10 iterations, each time refining the mask by excluding previously identified outliers and recomputing the smoothed flux and residuals. If the mask remained unchanged between successive iterations, the algorithm terminated early.
\end{enumerate}

\subsection{Two-stage Detrending with \text{\texttt{wotan}}} \label{sec:detrending}

After preparing the light curves, we performed a two-fold detrending process to remove both low- and high-frequency non-transit signals while preserving the strength of potential planetary transits.

\subsubsection{Primary Sinusoidal Signal Search (GLS)}

First, we identified the strongest sinusoidal signal in the sigma-clipped light curve using a generalized Lomb-Scargle (GLS) periodogram \citep{zechmeister_kurster_2009}. To mitigate the influence of common TESS systematics (e.g., momentum dumps), we performed this search \textit{after} outlier masking, which reduces sensitivity to low-frequency instrumental trends. We refer to the period corresponding to the highest-significance peak as $P_{\text{GLS, 1}}$. 

If \( P_{\text{GLS, 1}} < 1.0 \) day, we halved the primary detrending window to better preserve high-frequency variability. This adjustment aimed to improve detrending in light curves that display rapid, periodic structure.

\subsubsection{Primary Detrending}
We then detrended the light curve using \texttt{wotan} \citep{Wotan}, with the aim of removing the sinusoidal signal detected above in full or in part. We employed the built-in `cosine' method with a window length of 5 days (or 2.5 days for the aforementioned active targets). This step removed lower-frequency trends while preserving astrophysical signals such as rapid rotation signatures, flares, and transits. The effect of this primary detrending is shown in \autoref{fig:detrending_example}.

We then reapplied the iterative sigma-clipping process from Section \ref{sec:data_prep} to remove residual systematics and ensure a smooth baseline.

\subsubsection{Secondary Sinusoidal Signal Search (GLS)}
After the primary detrending, we computed a second GLS periodogram from the detrended light curve. We once again used the strongest GLS period ($P_\text{GLS, 2}$) to determine the correction to the secondary detrending window as before.

\subsubsection{Primary Transit Search (BLS)}
We searched for periodic, box-shaped transit signals in the single-detrended light curve using a Box-Least Squares (BLS) analysis \citep{kovacs_box-fitting_2002}. We used the \texttt{lightkurve.to\_periodogram} function \citep{lightkurve} to compute the BLS power spectrum by phase-folding the light curve across trial periods ranging from 1.0 to 12.0 days.\footnote{While our search for GEMS is restricted to orbital periods between 1.0 and 10.0 days, we extended the trial period range by two additional days to avoid edge effects and ensure robust detection near the upper boundary.}, using an oversampling factor of 7 to reduce the risk of missing potential transits. We defined the expected transit duration (the width of our test box) as 0.05 days, or 1.2 hours. This served only as an approximation, as we fit for the true transit duration later. This step yielded an initial estimate of the transit period ($P_\text{BLS, 1}$), the time of first transit ($T_\text{init, 1}$), and the approximate transit depth.

We also generated an ``Anti-BLS" periodogram using the inverse light curve (1/Flux), effectively searching for periodic brightening events instead of periodic dips. A match between the strongest detected periods in both the standard BLS and the ``Anti-BLS" periodograms (within 2\% of the BLS period) indicated that the signals were most likely sinusoidal and hence likely due to rotational modulation. This has been demonstrated to be more effective than trying to match the GLS period with that obtained from the BLS periodogram and has been used successfully in previous studies \citep[e.g.,][]{gan_occurrence_2023}. See Section \ref{sec:flags} for the quality flag associated with this check.

\subsubsection{Primary \texttt{batman} Fit}
We generated a transit model with \texttt{batman}, taking the $P_\text{BLS, 1}$, $T_\text{init, 1}$, and primary BLS transit depth at maximum power as priors for optimization. We also set the prior values for impact parameter $b$ and $a/R_*$ as 0.3 and 15.0 respectively. We assumed an inclination angle of $90\degree$, a circular orbit ($e=0$, $\omega_\star=90^\circ$), and quadratic limb-darkening coefficients from \cite{claret_limb_2017}. We fit the phase-folded data using \texttt{scipy.optimize.minimize} \citep{scipy} to refine the coarse period estimate ($P_\text{BLS, 1}$), initial transit time estimate ($T_\text{init, 1}$), and transit duration estimate. We optimized the BLS fit with the Nelder-Mead method and a convergence tolerance of $10^{-6}$ \citep{neldermead1965}. The Nelder-Mead method is particularly useful for fitting complex models without the need for gradients, but has the disadvantage of being relatively computationally expensive (see Section \ref{sec:injection}).

\subsubsection{Transit Masking and Secondary Detrending}
We used the estimated transit parameters identified in the previous step to temporarily mask all detected transits in the single-detrended light curve. We then detrended the remaining data using a finer window length of 0.5 days (or 0.25 days if $P_\text{GLS, 2}<1.0~\mathrm{day}$). Using the \texttt{numpy} package \citep{harris_array_2020}, we interpolated the masked regions with a one-dimensional linear interpolation, which effectively bridges short data gaps like those caused by transits. We subtracted the resulting trend from the single-detrended data to produce a twice-flattened light curve. \autoref{fig:detrending_example} shows an example of this step. While the single-detrended light curve retains some high-frequency modulation, the secondary detrending removes it cleanly and yields a satisfactorily flat curve that preserves the transit signal.

\subsubsection{Secondary Transit Search (BLS)} \label{sec:tessminer_fit}
We performed a second pair of BLS and Anti-BLS analyses to extract the period ($P_{\text{BLS, 2}}$), initial transit time ($T_\text{init, 2}$), and transit duration from the double-detrended light curve. 

\subsubsection{Secondary \texttt{batman} Fit}

We then fitted a second \texttt{batman} model with the same assumptions and methods as in the primary fit. However, this time, our priors were taken from the second BLS analysis. The best-fit parameters from this second fit ($P_\text{BLS, 2}$, $T_\text{init, 2}$, transit duration, radius ratio ($R_p / R_\star$), impact parameter ($b$), and scaled semi-major axis ($a/R_\star$)) were used to classify each system and assess whether a genuine transit signal was detected in the data.

\subsection{Target Classification} \label{sec:flags}
Our search criteria was $0.714~R_J < R_p < 1.5 R_J$ and periods spanning 1 -- 10 days.\footnote{The lower period bound of 1 day helps avoid contamination from short-timescale variability due to stellar activity or instrumental systematics, which are more prominent at high cadence. The upper bound of 10 days guarantees at least 2--3 transits for a planet observed over a full 27-day TESS sector. Ultra-short period and wide-separation GEMS are beyond the scope of this work.} We classified each observation as either a likely candidate (``Good Fit") or an unlikely candidate (``Poor Fit") based on several quality flags included in \texttt{TESS-miner}, described below:
\begin{itemize}
    \item \textbf{\texttt{Chi2Flag}:} this flag is \texttt{True} if the $\chi^2_{\text{red}}$ value of the transit fit (see Section \ref{sec:tessminer_fit}) is greater than 3.0 or less than 0.3 to prevent under- and over-fitting respectively.
    \item \textbf{\texttt{NoInTransitFlag}:} \texttt{True} if no in-transit data points are identified during the final transit search (see Section \ref{sec:tessminer_fit}). This occurs when the only detected periodicity corresponds to ``predicted" transits that fall entirely within data gaps, leaving no observable transit signals in the available data.
    \item \textbf{\texttt{DensityFlag}:} \texttt{True} if the fitted stellar density from Section~\ref{sec:tessminer_fit} falls outside specified, customizable bounds. Assuming a circular orbit or with knowledge of orbital eccentricity $e$, a genuine planetary transit provides an opportunity to measure the host star’s density \citep{seager_unique_2003, winn_exotransits_2010}.\footnote{We chose to fit circular models due to the short orbital periods being probed here.} In our full sample, the star possessed of the lowest density was TIC 307928780 ($3.12$ g\,cm$^{-3}$), and the star with the highest density was TIC 165086573 ($92.59$ g\,cm$^{-3}$). We adopted a lower limit of 1\,\gcmcubed{} and an upper limit of 200\,\gcmcubed{}, with the limits conservatively defined to exclude only the poorest fits.
    \item \textbf{\texttt{AntiTransitBLSFlag1}/\texttt{AntiTransitBLSFlag2}:} \texttt{True} if the period detected in either of the two ``Anti-BLS" periodograms coincides (within 2\%) with the BLS maximum power peak from Sections \ref{sec:detrending} and/or \ref{sec:tessminer_fit}. This flag aims to identify sinusoidal signals from stellar rotation which have nonetheless been identified as a BLS signal.
    \item \textbf{\texttt{RadiusFlag}:} \texttt{True} if the planetary radius fit in Section \ref{sec:tessminer_fit} is less than $6~R_{\oplus}$ ($\sim 0.5~R_J$). This serves to remove all non-giant planet candidates from consideration. While our definition of GEMS sets a lower bound at 0.7 $R_J$, we adopted a slightly smaller limit to account for fitting inaccuracy, grazing transit geometries, and up to 25\% dilution that may have remained uncorrected by the \texttt{tglc} pipeline.
\end{itemize}

If \textit{any} of the above flags were \texttt{True}, we classified the light curve in question as a ``Poor Fit." If all of the above flags were \texttt{False}, we classified the light curve as a ``Good Fit." If a target was observed in multiple TESS sectors during the PM and EM1, we advanced it to the vetting stage if at least one sector was classified as a ``Good Fit." This approach ensured that a system was not discarded solely due to contamination or noise in a single sector (e.g., from scattered light.)

\section{Candidate Vetting} \label{sec:vetting}
\subsection{Stage 0: Initial \texttt{TESS-miner} Result Breakdown}

\begin{figure}
    \centering
    \includegraphics[width=\linewidth]{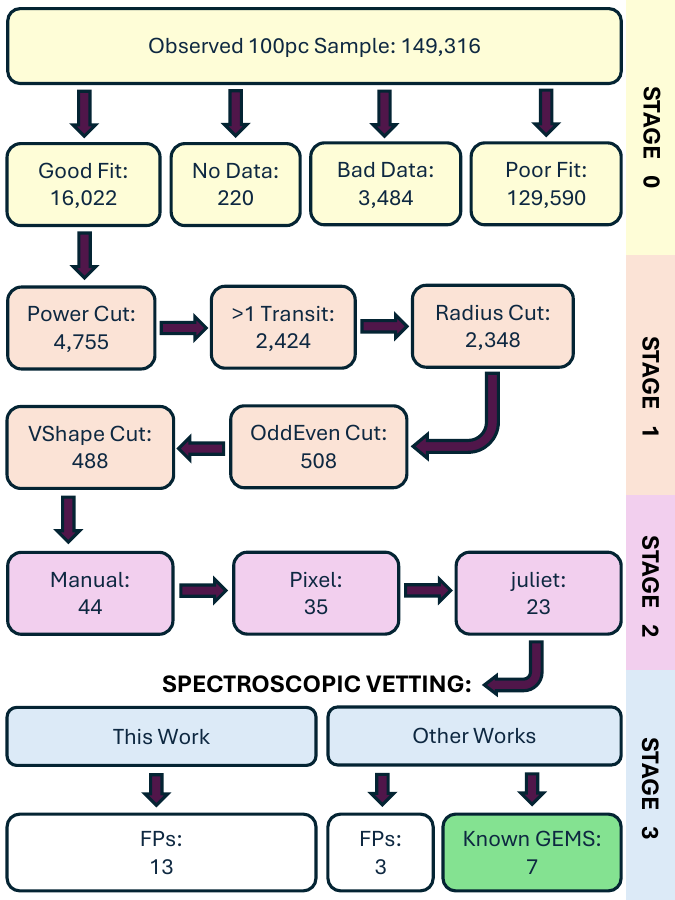}
    \caption{The cascade of candidates from the initial sample through our final GEMS list. Steps in our search include analysis with \texttt{TESS-miner} (Stage 0; yellow), auto-vetting (Stage 1; orange), by-eye and otherwise supervised inspection steps (Stage 2; purple), and spectroscopic validation (Stage 3; blue). Final results are divided into false positives (FPs; white) and previously confirmed GEMS (green).}
    \label{fig:flow}
\end{figure}

\begin{figure*}
    \centering
    \includegraphics[width=0.9\linewidth]{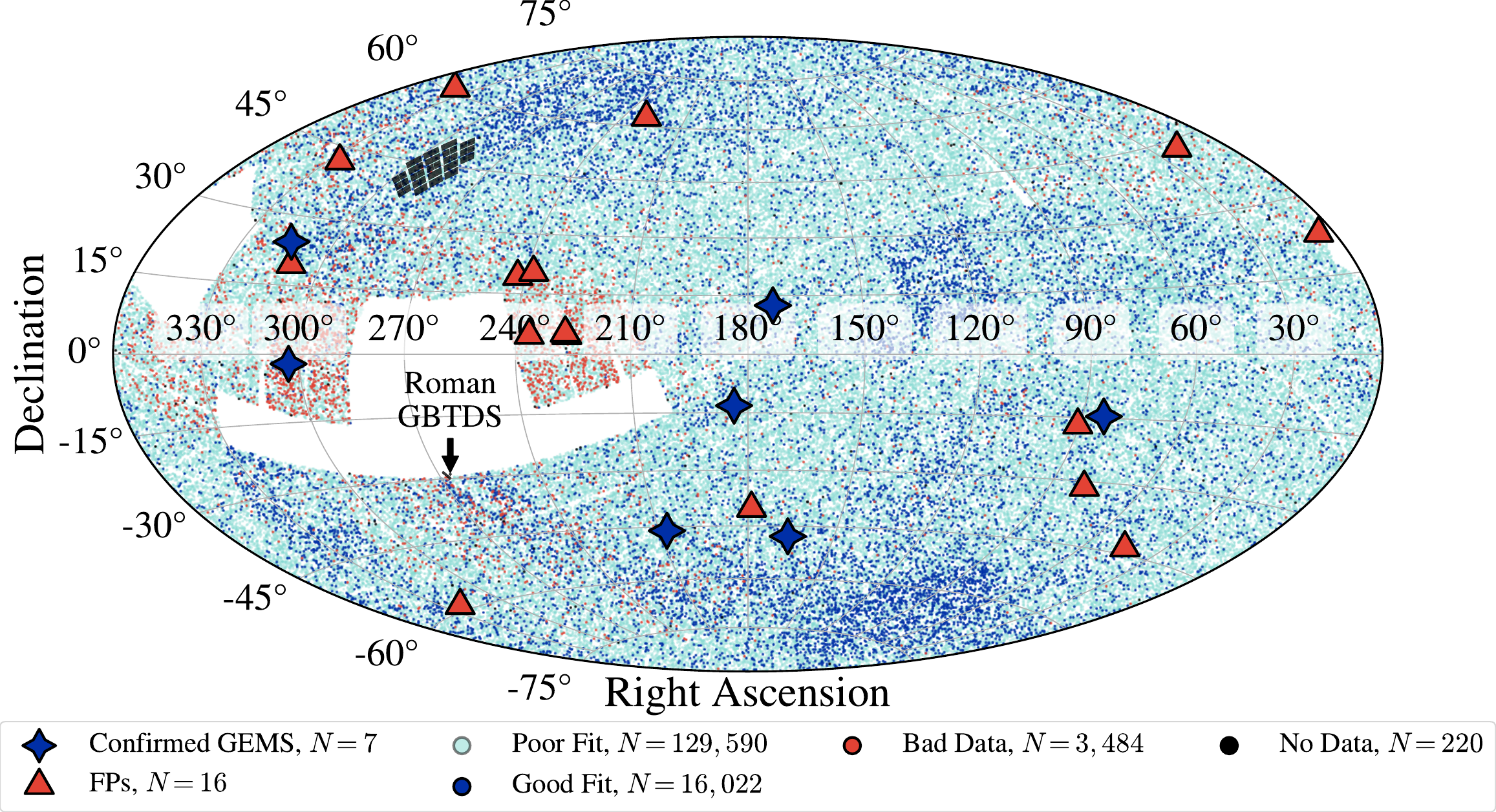}
    \caption{Map of all targets. Poor Fits are shown in pale turquoise, Good Fits in dark blue, Bad Data in red, and targets lacking TGLC data in black points.\footnote{Note that legend markers are enlarged for clarity. Black points indicate stars lacking TESS data and should not be confused with the black polygons denoting the Kepler and Roman survey footprints overlaid on the figure.} Spectroscopically vetted targets (\autoref{tab:lit_search}) are overplotted: FPs as red triangles and known GEMS as dark blue astroids. The FPs denoted here include those identified via spectroscopic followup by this work (13) and by previous works (3). The Kepler footprint \citep[black bars;][]{2016ascl.soft01009M} and Nancy Roman Galactic Bulge Time Domain Survey footprint\footnote{Adopted centers from \url{https://github.com/mtpenny/gbtds_optimizer/blob/main/field_layouts/layout_40395.centers}} (black polygons indicated by arrow; \citet{Spergel15, Penny2019,rotac2025}) are shown for reference.}
    \label{fig:surveymap}
\end{figure*}

We processed our initial sample of 149,316 unique stars with \texttt{TESS-miner}, resulting in the breakdown shown in \autoref{fig:flow}. Less than $0.15\%$ of the M-dwarfs in our sample lacked associated \texttt{tglc} light curves, preventing further analysis. This left 149,069 unique stars with 477,999 sectors of data for our analysis. Of these, \texttt{TESS-miner} flagged around $2.3\%$ as having ``Bad Data;" $86.9\%$ as ``Poor Fits"; and $10.7\%$ as ``Good Fits." The 16,022 stars classified as ``Good Fits" had been observed across 86,404 single-sector light curves, giving an average of 5.39 sectors per star. We refer hereafter to the initial \texttt{TESS-miner} search and its result breakdown as ``Stage 0."

\autoref{fig:surveymap} shows a map of our full sample, colored by \texttt{TESS-miner} classification. ``Bad Data" (red points) cluster around the galactic bulge, where stellar crowding contaminates flux; ``Good Fits" (dark blue points) concentrate near the poles and ecliptic, where TESS coverage is denser. Meanwhile, all targets that survive to the spectroscopic vetting stage are also included on the map (see Section \ref{sec:specvetting} for details; objects are enumerated in \autoref{tab:lit_search}).

\subsection{Stage 1: Auto-vetting Steps}
\label{sec:autovetting}
We then performed a series of auto-vetting steps to sort through the 16,022 stars classified as ``Good Fits" during Stage 0. We chose all thresholds used for target classification within \texttt{TESS-miner} itself (Section \ref{sec:flags}) and in our auto-vetting steps to retain all GEMS within 200 pc (see Section \ref{sec:sample}.) We also included generous margins to account for poor data quality and faint targets. The cuts were as follows:

\begin{enumerate}
    \item \textbf{BLS Power Cut:} We removed all objects from consideration whose maximum BLS power was below 100. We use BLS power here instead of traditional Gaussian S/N because it is specifically designed to detect periodic, box-shaped transits by phase-folding light curves. This makes it more sensitive and robust to real transit signals in noisy, irregular data. The formulation for BLS power is defined by Equation 11 in \citet{kovacs_box-fitting_2002} as the square of the ``effective S/N" $\alpha=\frac{\delta}{\sigma}\sqrt{nq}$, where $\delta$ is the transit depth, $\sigma$ is the per-point noise, $n$ is the total number of data points, and $q$ is the fractional transit duration.
    
    \item \textbf{Transit Number Cut:} We then excluded all objects with $< 2$ detected transits per ``Good Fit" sector. Our search spans periods from 1 to 10 days, and since each TESS sector covers approximately 27 days, the longest-period planets within our range are expected to exhibit at least two transits per light curve.
    
    \item \textbf{Radius Cut:} We excluded all candidates with fitted planetary radii greater than 22 $R_{\oplus}$ ($2.0~R_J$). We chose this threshold to remove likely EBs from our sample. Radius inflation beyond $2~R_J$ is unlikely for GEMS, as M-dwarfs do not emit enough radiation to trigger the same inflation mechanisms observed in hot Jupiters. At the same time, our threshold allows for some margin beyond our nominal upper bound of 1.5 $R_J$.
    \item \textbf{OddEven Cut:} \texttt{TESS-miner} includes several built-in vetting flags, one of which is the \texttt{OddEven} flag. This flag compares the mean depths of odd and even transits and is triggered when the difference exceeds 25\%. We designed this flag to identify EB false positives in cases where (a) the binary orbit is circular, and (b) the BLS algorithm has locked onto half the true orbital period, causing both primary and secondary eclipses to be interpreted as transits. In such cases, the differing depths of odd and even events reflect the luminosity contrast between the binary components.
    \item \textbf{V-Shaped Cut:} We defined an additional cut on V-shaped transits using the same method as appears in the \texttt{LEO-vetter} auto-vetting package \citep{Kunimoto_2024}. If $b + R_p/R_\star>1.5$ where $b$ is the fitted impact parameter, we classified the light curve as V-shaped and removed it from consideration. A similar metric was used in the Kepler mission \citep{kepdr25} to reliably identify EBs as too deep (large $R_p / R_\star$) or extremely grazing (large $b$). This threshold is the default choice for \texttt{LEO-vetter}, and preserved our test sample of all known GEMS within 200 pc. We found that 24.3\% of surviving candidates had fitted impact parameters $b > 0.85$ even after applying this cut, alleviating concerns that it might exclude grazing planets from our search.
\end{enumerate}

We refer to these auto-vetting steps collectively as ``Stage 1" of analysis. Following Stage 1, 488 candidates remained with 709 sectors of data. 

\subsection{Stage 2a: Manual Vetting} \label{manual}

\begin{figure}
    \centering
    \includegraphics[width=\linewidth]{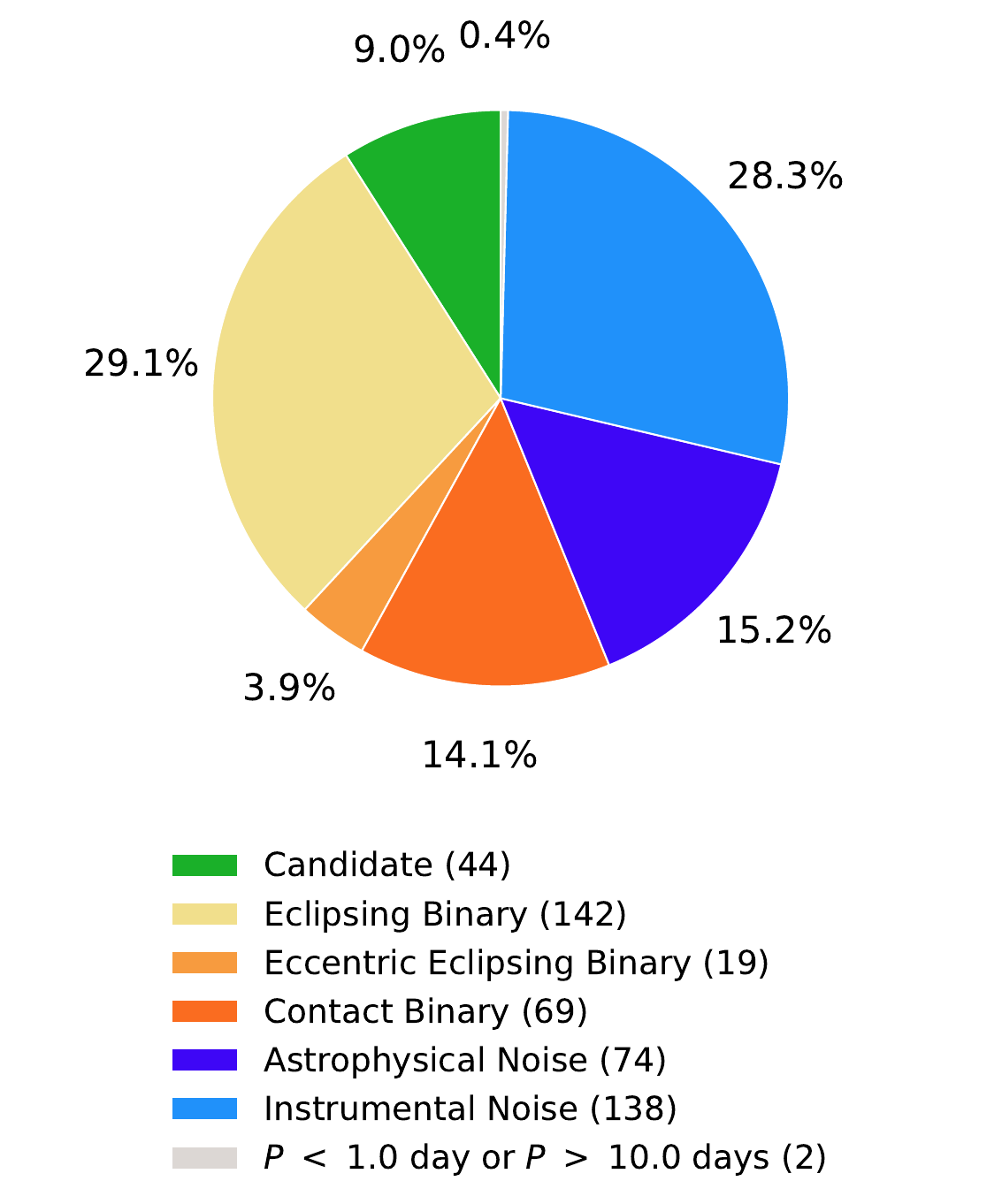}
    \caption{A breakdown of the types of signals detected by \texttt{TESS-miner} within the 488 candidates passed to Stage 2a of analysis (manual vetting). These include valid candidates, EBs, eccentric EBs, contact binaries, instrumental noise, and astrophysical noise. The latter comprises host and background activity signals. Finally, a small sliver of our FPs were excluded based on having a true orbital period outside the scope of this work.}
    \label{fig:manual_pie}
\end{figure}

We manually inspected each of the surviving 488 candidates and divided them into the following categories:

\begin{itemize}
    \item \textbf{FP due to instrumental noise:} No clear transit-like feature. In some cases, a cluster of correlated noise can appear in a TESS light curve, oftentimes near the beginning or end of a sector or adjacent the mid-sector data gap. These artifacts are typically caused by instrumental effects such as thermal settling and momentum dumps while the telescope reorients. Although we designed our edge-clipping procedure (see Section \ref{sec:data_prep}) to reduce the impact of these artifacts, certain noise signatures were still flagged by the BLS algorithm as a transit.
    \item \textbf{FP due to astrophysical noise:} Instances where the BLS search detected signals caused by strong stellar activity or rotational modulation, either from the host star or a background object (i.e., a BY Draconis variable). Signals from background sources are typically identifiable by their appearance in only one of several observed sectors, often in combination with a crowded field.
    \item \textbf{FP due to EB:} Instances where the identified signal had an obvious difference in odd and even transit depths. For these signals, \texttt{TESS-miner} either identified the correct period, and found that the difference between odd and even transit depths was less than 25\%, therefore not triggering the \texttt{OddEvenFlag}; or had identified an alias of two or four times the correct period (i.e., only the odd or even transits), thereby invalidating said flag. 
    \item \textbf{FP due to an eccentric EB:} A handful of objects were eccentric EB systems, where the secondary transit signal did not occur at the exact midpoint between primary transits. \texttt{TESS-miner} does not yet account for this type of signal pattern.\footnote{We note here that M-dwarf/M-dwarf binaries, of which these objects are prime examples, tend to have higher primary to secondary mass ratios, as well as lower eccentricities, than other EB spectral pairings \citep{Fontanive2018, dupuy_individual_2017}. This makes eccentric EBs quite rare in our sample, as well as interesting targets for future research in that field.}
    \item \textbf{FP due to contact binary:} Contact binaries display constant flux modulation due to the distortion of the component stars \citep{Mattei1999}, and periods of less than 0.7 days \citep{Hilditch2001}.\footnote{In these cases, \texttt{TESS-miner} identified an alias of the true period which was within the search bounds of 1.0 to 12.0 days.} Their signal often includes equal primary and secondary transit depths due to roughly equivalent component radii \citep{Kuiper1941, Lucy1967}. These features combine to produce a characteristic scalloped pattern.
    \item \textbf{Excluded from our search due to orbital period:} instances where the identified signal appeared to be an alias of a true period $<1.0$ day or $>10.0$ days. 
    \item \textbf{Possible candidates:} all remaining objects.
\end{itemize}

The breakdown of these FP categories can be seen in \autoref{fig:manual_pie}. The bulk of FPs were caused by a combination of different binary system morphologies and instrumental noise. Examples of each category can be seen in Appendix \autoref{fig:man_ex}. Removing these FPs resulted in 44 remaining candidates. 

\subsection{Stage 2b: Pixel-Pixel Maps}\label{sec:pixelpixel}
We subjected the surviving 44 candidates to visual pixel-level vetting using the \texttt{tglc} package. The \texttt{tglc} pipeline constructs de-blended light curves from FFI cutouts by modeling and subtracting flux contamination from nearby stars, using Gaia-based positional and photometric priors (Han et al. 2025, in prep). This yields photometry that reflects the FFI cutout flux at the per-pixel level. If the target star is the true source of the transit signal, the normalized light curves from pixels centered on its coordinates will show consistent transit depths. In contrast, if the signal originates from a nearby contaminant, the offset PSF will cause the transit depth to vary across the pixel-level light curves. We generated per-pixel light curve maps for each observed sector per each of the 44 candidates. We excluded nine candidates in which the transit signal identified by \texttt{TESS-miner} appeared strongest in pixels centered on a different star than the presumed host. Examples of (a) a clear pass and (b) a clear failure in this step are included in Appendix \autoref{fig:pxpx}. In four cases, crowded fields prevented a confident assessment, and these candidates continued to the next stage of vetting alongside the remaining survivors.

\subsection{Stage 2c: Stellar Density Estimation with \texttt{juliet}}\label{sec:juliet}

We used the Bayesian modeling tool \texttt{juliet} \citep{espinoza_juliet_2019} to fit the double-detrended light curves from \texttt{TESS-miner} for the 35 candidates that passed pixel-level vetting. \texttt{juliet} employs dynamic nested sampling via the \texttt{dynesty} package \citep{speagle_dynesty_2020} to sample the full posterior distribution. Unlike the Nelder-Mead optimizer used for preliminary fits in \texttt{TESS-miner}, \texttt{juliet} supports informative priors and explicit modeling of systematics, enabling more robust discrimination between planetary transits and EBs.

\begin{figure*}[ht]
    \centering
    \includegraphics[width=0.9\textwidth]{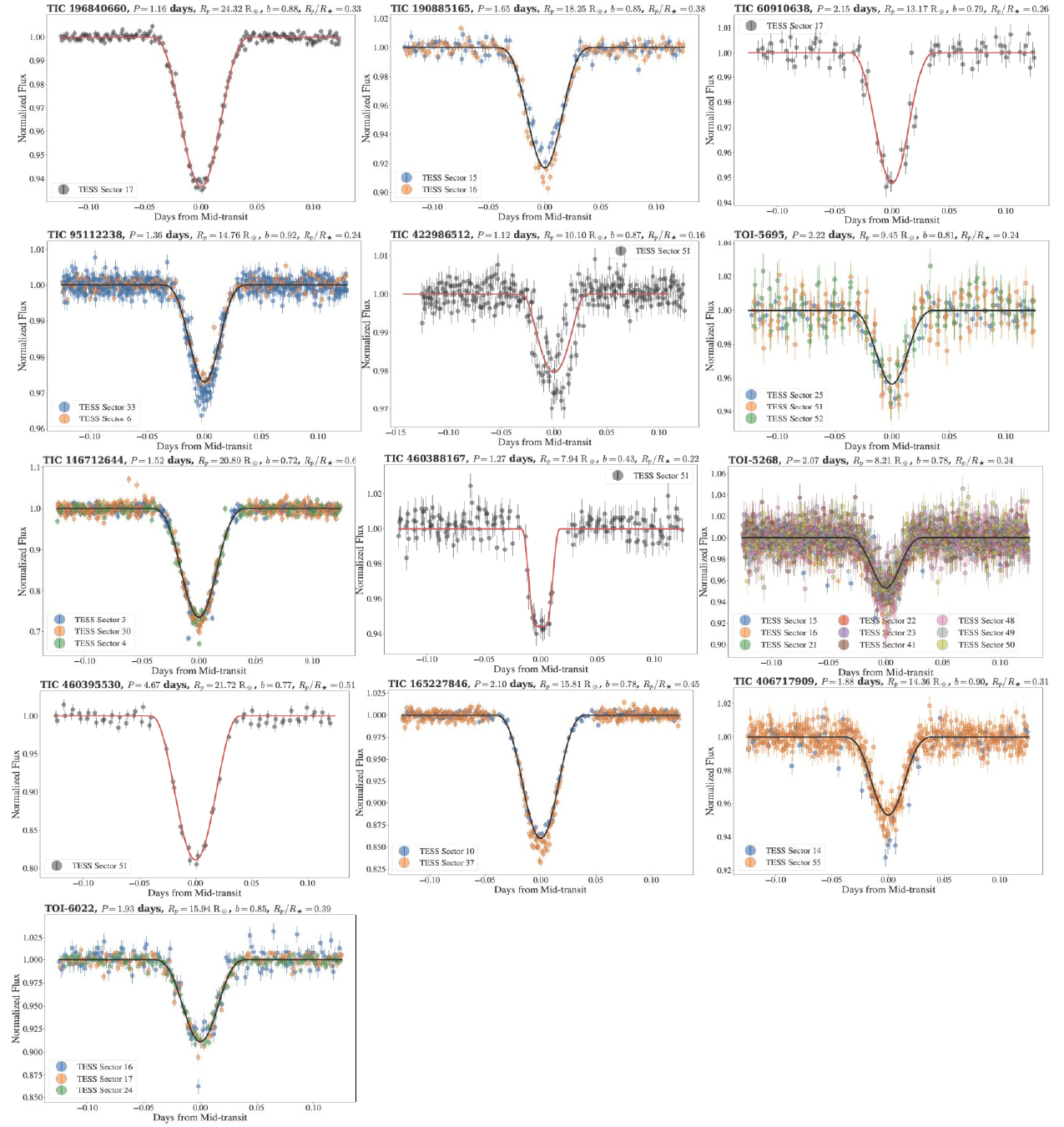}
    \caption{\texttt{juliet} fits for the 13 TICs cited in \autoref{tab:lit_search} as new astrophysical false positives by this work.}
    \label{fig:julietfits}
\end{figure*}

For this work, we used the default \texttt{dynesty} convergence criterion of $\Delta\ln Z < 0.01$. We generated transit models using \texttt{batman} \citep{kreidberg_batman_2015}, adopting a quadratic limb-darkening law and sampling coefficients from uniform priors following \citet{kipping_efficient_2013}. To account for finite integration times, we applied a supersampling factor of 30. We placed Gaussian priors ($\sigma = 0.1~\mathrm{days}$) on the period and mid-transit time derived from \texttt{TESS-miner}. All fits assumed circular orbits ($e = 0$, $\omega_\star = 90^\circ$) and a fixed dilution factor of unity.\footnote{Although GEMS may retain spin-orbit misalignment and eccentricity (Section \ref{sec:intro}), only one confirmed GEM to date demonstrates appreciable eccentricity (TOI-6330 b; \citet{hotnisky_2025_searching}), and it does not lie within 100 pc.} We fixed the dilution factor because TESS photometry alone (i.e., without high-resolution imaging or multi-band photometry) cannot constrain contamination. The assumption of zero eccentricity yields smaller radius estimates and is a conservative choice when modeling low-precision TESS data \citep[e.g.,][]{Kipping2008}. The photometric model included a sector-specific white-noise jitter term added in quadrature to the reported uncertainties. We jointly fitted all available TESS sectors per target, allowing sector-specific baselines and jitter terms while sharing limb-darkening parameters. The final fits and phase-folded light curves appear in \autoref{fig:julietfits}.

We used the posterior distributions from the \texttt{juliet} fits to identify FPs. First, we excluded candidates whose fitted planet radii fell more than $1\sigma$ below $0.7~\mathrm{R_J}$, outside the scope of our search. We also flagged systems as likely background eclipsing binaries (BEBs) if the stellar density inferred from the transit shape differed from the expected density (see Section~\ref{sec:sample}) by more than $5\sigma$. Such discrepancies suggested that the fitted transit signals do not originate from the target star. 

The \texttt{juliet} fitting stage served a different purpose than the \texttt{TESS-miner} \texttt{DensityFlag} (see Section~\ref{sec:flags}). Whereas the \texttt{DensityFlag} applied broad thresholds to eliminate clearly unphysical fits (e.g., noise-driven BLS detections), the \texttt{juliet}-based density check specifically targeted BEBs. This step removed 12 FPs, leaving 23 objects of interest for spectroscopic vetting.

\subsection{Stage 3: Observational and Literature Vetting}
\label{sec:specvetting}
\begin{deluxetable*}{|p{2.0cm}|p{2.9cm}|p{2.9cm}|p{2.5cm}|p{1.5cm}|p{3.8cm}|}
\tablecaption{The 23 objects which survived all auto and manual vetting prior to spectroscopic and literature validation. Of those 23, 7 are previously confirmed GEMS and 16 are astrophysical false positives (i.e., not transiting planets.) SB1/2/3 refer to single, double, or triple spectroscopic binaries.\label{tab:lit_search}}
\tabletypesize{\normalsize}
\tablewidth{\textwidth}
\tablehead{
\hline
\textbf{TIC} & \textbf{Companion} & \textbf{Additional Data} & \textbf{Status} & \textbf{Spectral Type} & \textbf{Reference}
}
\startdata
\multicolumn{6}{c}{\textbf{Confirmed}} \\
\hline
243921117 & WASP 80 b & -- & Confirmed & Early & \citet{triaud_wasp_80b_2013} \\
46432937 & TIC 46432937 b & -- & Confirmed & Early & \citet{hartman_toi_762_2024} \\
243641947 & TOI-3235 b & -- & Confirmed & Mid & \citet{hobson_toi-3235_2023} \\
335590096 & TOI-4860 b & -- & Confirmed & Mid & \citet{almenara_toi-4860_2023} \\
419411415 & TOI-5205 b & -- & Confirmed & Mid & \citet{kanodia_toi-5205b_2023} \\
178709444 & TOI-762 A b & -- & Confirmed & Mid & \citet{hartman_toi_762_2024} \\
67512645 & TOI-6894 b & -- & Confirmed & Late & \citet{Bryant2025} \\
\hline
\multicolumn{6}{c}{\textbf{False Positives}} \\
\hline
196840660 & -- & HPF & SB3 & Early & This work \\
190885165 & -- & HPF & SB2 & Early & This work \\
60910638 & -- & NEID & SB2 & Early & This work \\
95112238 & -- & NEID & SB2 & Early & This work \\
422986512 & -- & HPF & SB2 & Early & This work \\
103865797 & TOI-5695.01 & HPF & SB2 & Mid & This work \\
77951245 & TOI-450.01 & -- & EB & Mid & \citet{Prsa2022} \\
146712644 & -- & -- & EB & Mid & This work \\
460388167 & -- & HPF & SB1 & Mid & This work \\
202468443 & TOI-5268.01 & HPF & SB3 & Mid & This work \\
460395530 & -- & HPF & SB2 & Mid & This work \\
165227846 & -- & WINERED & SB2 & Mid & This work \\
406717909 & -- & HPF & SB2 & Mid & This work \\
455947620 & TOI-6022.01 & -- & EB & Mid & This work \\
272519426 & TOI-5693.01 & -- & BD & Mid & \citet{irwin_nltt_2010} \\
381856446 & WASP 145 A b  (TIC 381856447 b) & -- & Background Planet & Mid & \citet{hellier_2019_wasp145ab} \\
\enddata
\end{deluxetable*}

At this point, we referred back to our sample of known GEMS within 100 pc (see Section \ref{sec:sample}). We confirmed that we had successfully recovered all seven objects, reinforcing the robustness of our pipeline (see \autoref{tab:lit_search}). Furthermore, we confirmed that we had recovered or reasonably discounted from our analysis all TOIs occurring within our input catalog and search bounds as of August 10, 2025. Only one TOI (5850, a spectroscopically confirmed single-lined binary),  was removed prior to Stage 2, due to noisy observational data. This case, alongside other TOIs not explicitly mentioned in the main text of this work, is fully explained in Appendix \autoref{tab:tois}.

We then consulted the TESS EB catalog \citep{Prsa2022} as well as previous GEMS and EB surveys to determine the disposition of our remaining 16 candidates. These objects are listed in the bottom section of \autoref{tab:lit_search}. We identified one candidate as an EB \citep{Prsa2022} and one candidate as a BD \citep{irwin_nltt_2010}. We identified a further signal, ostensibly from TIC 381856446 / WASP 145 B, as a confirmed hot Jupiter orbiting the target's much brighter K2 V companion, WASP 145 A \citep{hellier_2019_wasp145ab}. The remaining 13 candidates were subjected to spectroscopic validation using HPF, NEID and WINERED (see below for details on the use of these instruments.) 

\textbf{Spectroscopic vetting removed all new GEMS candidates, reducing our total GEMS sample from 20 to 7. This reduction of 65\% highlights the importance of spectroscopic follow-up to eliminate FPs. 
}

\renewcommand{\arraystretch}{1.5} 
\begin{deluxetable*}{c|c|c|c|c|c|c}[ht!]
\tablewidth{\textwidth}
\tablecaption{Selected orbital parameters of all confirmed GEMS retrieved in our sample. Object parameters are quoted from the references in the final column. All stellar radii are taken from \citet{Mann15}. 
Objects are sorted by $R_p$. \label{tab:finalsample}}
\tablehead{
\colhead{\textbf{TIC}} & \colhead{\textbf{Companion}} & \colhead{\textbf{Period (days)}} &
\colhead{$\mathbf{R_p ~(R_J)}$} & \colhead{\textbf{Impact Param.}} & 
\colhead{$\mathbf{R_\star ~(R_\odot)}$} & \colhead{\textbf{Reference}}
}
\startdata
178709444 & TOI-762 A b & $3.47\pm7.2\cdot10^{-7}$ & $0.744\pm0.017$ & $0.7556^{+0.0057}_{-0.0067}$ & $0.428$ & \citet{hartman_toi_762_2024} \\
335590096 & TOI-4860 b & $1.52\pm3.0\cdot10^{-7}$ & $0.77\pm0.03$ & $0.29^{+0.06}_{-0.11}$ & $0.355$ & \citet{almenara_toi-4860_2023} \\
243921117 & WASP 80 b & $3.07^{+8.3\cdot10^{-7}}_{-7.9\cdot10^{-7}}$ & $0.999^{+0.030}_{-0.031}$ & $0.215^{+0.020}_{-0.022}$ & $0.636$ & \citet{triaud_wasp_80b_2013} \\
243641947 & TOI-3235 b & $2.59\pm4.1\cdot10^{-7}$ & $1.02\pm0.044$ & $0.511^{+0.011}_{-0.012}$ & $0.371$ & \citet{hobson_toi-3235_2023} \\
419411415 & TOI-5205 b & $1.63\pm1.0\cdot10^{-6}$ & $1.03 \pm 0.03$ & $0.0016 \pm 0.0002$ & $0.400$ & \citet{kanodia_toi-5205b_2023} \\
46432937 & TIC 46432937 b & $1.44\pm8.7\cdot10^{-8}$ & $1.19\pm0.030$ & $0.825^{+0.0157}_{-0.0073}$ & $0.539$ & \citet{hartman_toi_762_2024} \\
67512645 & TOI-6894 b & $3.37\pm0.000296$ & $1.33^{+0.57}_{-0.43}$ & $0.927^{+0.551}_{-0.530}$ & $0.220$ & \citet{Bryant2025} \\
\enddata
\end{deluxetable*}

\subsubsection{HPF -- Northern Hemisphere Targets I}

For M-dwarfs in the northern hemisphere, we used the Habitable-zone Planet Finder \citep[HPF; ][]{mahadevan_habitable-zone_2012, mahadevan_habitable-zone_2014} to obtain observations on our surviving candidates after the aforementioned checks.  HPF is a near-infrared (\(8080-12780\)\ \AA), environmentally stabilized \citep{stefansson_versatile_16}, fiber-fed \citep{kanodia_overview_2018} spectrograph with laser-frequency comb calibration \citep{Metcalf2019} on the 10 m Hobby-Eberly Telescope\footnote{ Based on observations obtained with the Hobby-Eberly Telescope (HET), which is a joint project of the University of Texas at Austin, the Pennsylvania State University, Ludwig-Ludwig-Maximilians-Universität München, and Georg-August Universität Gottingen. The HET is named in honor of its principal benefactors, William P. Hobby and Robert E. Eberly.} \citep[HET;][]{ramsey_early_1998, hill_current_2012,hill_hetdex_2021} at the McDonald Observatory in West Texas, USA. We used the \texttt{HxRGproc} package \citep{ninan_habitable-zone_2018} to correct for bias, non-linearity, cosmic rays and process the HPF slope images. We then used \texttt{barycorrpy} \citep{kanodia_python_2018} to perform the barycentric correction on the individual spectra, which is the Python implementation  of the algorithms from \cite{wright_barycentric_2014}. Each visit consisted of two exposures of 969 seconds each that are subsequently combined by weighted averaging. 

\begin{figure*}[ht]
    \centering
    \includegraphics[width=0.9\textwidth]{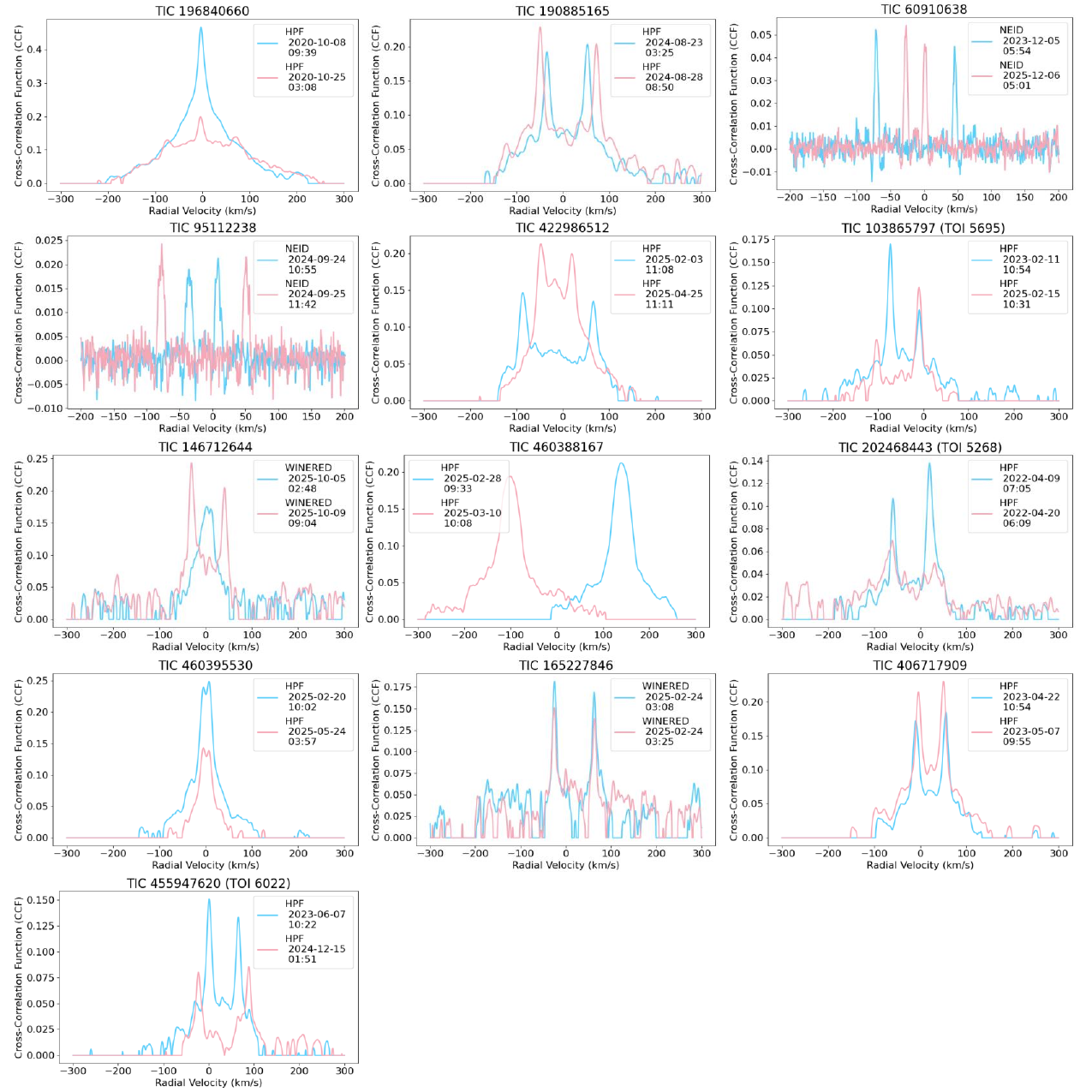}
    \caption{CCF plots for the 13 TICs cited in \autoref{tab:lit_search} as astrophysical false positives by this work. Objects are arranged in the same order as \autoref{fig:julietfits}. The different colors represent distinct spectroscopic epochs, with CCF shifts indicating Doppler motion inconsistent with a planetary companion. In particular, the presence of multiple peaks or large velocity variations between observations reveal the target as an EB rather than a single-lined star hosting a planet.}
    \label{fig:ccfs}
\end{figure*}

We followed the procedure described in \cite{Canas2023apogee} to cross-correlate the observed spectra with synthetic spectra from the PHOENIX library \citep{husser13} using the algorithms presented in \cite{Zucker2003}. We only cross-correlate the HPF spectra in the echelle orders $4-6$ ($\lambda\sim8530-8890$ \AA{}) and $14-18$ ($\lambda\sim9930-10760$ \AA{}) because of the minimal amount of telluric contamination in those regions. The resultant cross-correlation functions (CCFs) with HPF spectra are shown in \autoref{fig:ccfs}, and objects with multiple peaks are identified as FPs (hierarchical stellar systems). Using HPF spectra, we identified eight candidates that are spectroscopic binaries and are listed in \autoref{tab:lit_search}. A detailed characterization of these false positives is beyond the scope of this manuscript and is left to future work \citep[e.g.,][]{Boone2025}.

\subsubsection{NEID -- Northern Hemisphere Targets II} \label{neid}
We also used NEID, an environmentally stabilized \citep{robertson_ultrastable_2019}, fiber-fed \citep{kanodia_stable_2023} high-resolution optical (3800 -- 9300 \AA) spectrograph \citep{schwab_design_2016} on the 3.5~m WIYN telescope\footnote{The WIYN Observatory is a joint facility of the NSF’s National Optical-Infrared Astronomy Research Laboratory, Indiana University, the University of Wisconsin-Madison, Pennsylvania State University, Purdue University and Princeton University.} at Kitt Peak, Arizona, USA to spectroscopically validate some of our candidates. Each visit consisted of an 1800 s exposure using high-resolution mode ($R \sim$ 110,000). We used the Level 2 products from the NEID DRPv1.4.0\footnote{\url{https://neid.ipac.caltech.edu/docs/NEID-DRP/}}, which includes the wavelength-calibrated spectra. The NEID DRP also calculates the CCF for all objects by cross-correlating the observed spectra with binary masks \citep[e.g.,][]{baranne_elodie_1996} that are customized to various spectral types \citep{Bender2022}. We used these CCFs to confirm that TIC-95112238 and TIC-60910638 were both spectroscopic double-lined binaries (SB2) with two peaks in the CCF for each star (see \autoref{fig:ccfs}).

\subsubsection{WINERED -- Southern Hemisphere Targets} \label{winered}
WINERED (Warm INfrared Echelle spectrograph to Realize Extreme Dispersion and sensitivity) is a high-throughput high-resolution spectrograph \citep{ikeda_highly_2022} on the 6.5 m Magellan telescope at Las Campanas Observatory, Chile. For the southern hemisphere targets, we used WINERED in the \textit{Y} band in the HIRES mode ($R~\sim$ 68,000) with a 100 $\mu$m slit. We used the ABBA slit nodding with 300 s exposures each (4x 300 s = 1200 s total) to obtain simultaneous sky and science traces on the detector. We used the \texttt{WARP} pipeline \citep{2024PASP..136a4504H} to extract and reduce the spectra. We used the peak-finding algorithm in \texttt{scipy} \citep{scipy} to identify OH sky emission lines in the sky trace, selecting only those with widths greater than one resolution element (i.e., wavelength divided by spectral resolution). We created a binary mask at the position of these lines and broaden it by six resolution elements to include the line wings. The sky lines identified in this manner were then used to track the instrument velocity shifts over time with respect to a fiducial spectrum on an order-by-order basis by minimizing the $\chi^2$ in velocity space. We then fit a linear trend to the instrument drift across wavelength space based on the median wavelength of each order, and used this to assign a drift correction to each order. The order-by-order scatter across this linear trend was $\sim$ 100 -- 300 \ms{}, while the intra-night instrument drift was $\sim$ 5 \kms{}. Using the same algorithm as with the HPF spectra, we cross-correlated the drift-corrected spectra with a PHOENIX model using echelle orders $172-165$ ($\lambda\sim10230-10730$ \AA{}) to search for signs of binarity. We identified TIC-165227846 and TIC-243222192 as SB2s with the CCFs shown in \autoref{fig:ccfs}.

\section{Calculation of Occurrence Rates} \label{sec:overall_occ}

\subsection{Injection and Recovery}
\label{sec:injection}

\begin{deluxetable*}{c|c|c|c}
\setlength{\tabcolsep}{10pt}   
\tabletypesize{\normalsize}
\tablewidth{\linewidth}
\tablecaption{Injection population subsets by spectral type, TESS observation mission (PM: Primary Mission; EM1: First Extended Mission), and number of light curves per bin. Each light curve was injected with 1,000 unique planetary transit models.
\label{tab:inj_set}}
\tablehead{
    \colhead{\textbf{Type}} & 
    \colhead{\boldmath$M_{K_S}$} & 
    \colhead{\textbf{Mission}} & 
    \colhead{\boldmath$N_\text{LC}$}
}
\startdata
\hline
Early & $M_{K_S} < 6.0$          & PM & 7,404  \\
Early & $M_{K_S} < 6.0$          & EM1 & 8,307  \\
Mid   & $6.0 < M_{K_S} < 7.1$    & PM & 10,445 \\
Mid   & $6.0 < M_{K_S} < 7.1$    & EM1 & 10,512 \\
Late  & $M_{K_S} > 7.1$          & PM & 19,360 \\
Late  & $M_{K_S} > 7.1$          & EM1 & 16,281 \\
\hline
\enddata
\end{deluxetable*}

To determine the detection sensitivity of our pipeline, we randomly selected $\sim 20\%$ of observations initially classified by \texttt{TESS-miner} as unlikely candidates (``Poor Fits'') as shown in \autoref{tab:inj_set}. Accounting for occasional errors and runtime budgeting, we ultimately sampled 11,005 early-type stars, 14,466 mid-type stars, and 25,556 late-type stars. We then generated 1,000 injection parameter sets per star, each from a uniform distribution within the following bounds:

\begin{itemize}[itemsep=2pt, topsep=6pt, label=\raisebox{0.25ex}{\small\textbullet}]
    \item We injected planets with periods in the range $1.0 < P_{\text{inj}} < 10.0$ days. Although injections were limited to this range, the formal BLS search upper limit was 12 days to buffer against edge effects near the maximum period. For each injection, we computed $a/R_\star$ using the stellar density and the injected period.
    \item 0.7 $R_J$ $< R_{\text{p, inj}} < 1.5 ~R_J$, sampled in $R_p$ space.\footnote{We chose to sample injections in $R_p$ as opposed to $R_p/R_\star$ space because the variation in stellar radius within each spectral subtype is small enough that using $R_p/R_\star$ would introduce unnecessary redundancy.} Some GEMS fall near the edges of our injection bins, and we factor this in by bootstrapping across their radius and period uncertainties while estimating occurrence rates.
    \item $0 < b_{\text{inj}} < 1 + \frac{R_{\text{p, inj}}}{R_\star}$, sampled uniformly for each $\frac{R_{\text{p, inj}}}{R_\star}$.  This takes into account the full range of possible transit signals, including grazing signals. 
    \item We sampled $T_0$ randomly using a uniform prior where the limits were the minimum and maximum dates of the light curve.
\end{itemize}

We also separated TESS observations from PM and EM1 to evaluate how changes in observing cadence influenced transit recovery. Since all of our data are drawn from FFIs, the cadence differs between the missions: 30-minute cadence in PM compared to a 10-minute cadence in EM1. By comparing recovery rates across these categories, we can determine how cadence and stellar subtype affect detectability.

We added an \texttt{InjectRecoveryMode} mode in \texttt{TESS-miner} to create and inject \texttt{batman} models corresponding to each unique set of transit parameters for each light curve, assuming a circular orbit. We injected each light curve with each unique model prior to detrending. Each injected light curve was then processed through the steps detailed in Section \ref{sec:pipeline_overview}, as well as the vetting steps described in Section \ref{sec:vetting} (except for the \texttt{OddEven} tests). The \texttt{OddEven} cut was excluded from our injection/recovery vetting process in order to conserve computational resources. However, since none of the seven confirmed GEMS within our sample space triggered the \texttt{OddEvenFlag}, we did not expect its exclusion to result in inflated completeness or occurrence rates. In all, we performed $\sim 72$ million injections during this process to accurately characterize our pipeline detection sensitivity as it varies across spectral subtype (and hence brightness), TESS mission (PM vs EM1), and planet radius and orbital period. 

\subsection{Detection Sensitivity Grids} \label{sec:efficiency}

\begin{figure*}[ht]
    \centering

    \begin{subfigure}[t]{0.30\textwidth}
        \centering
        \includegraphics[width=\textwidth]{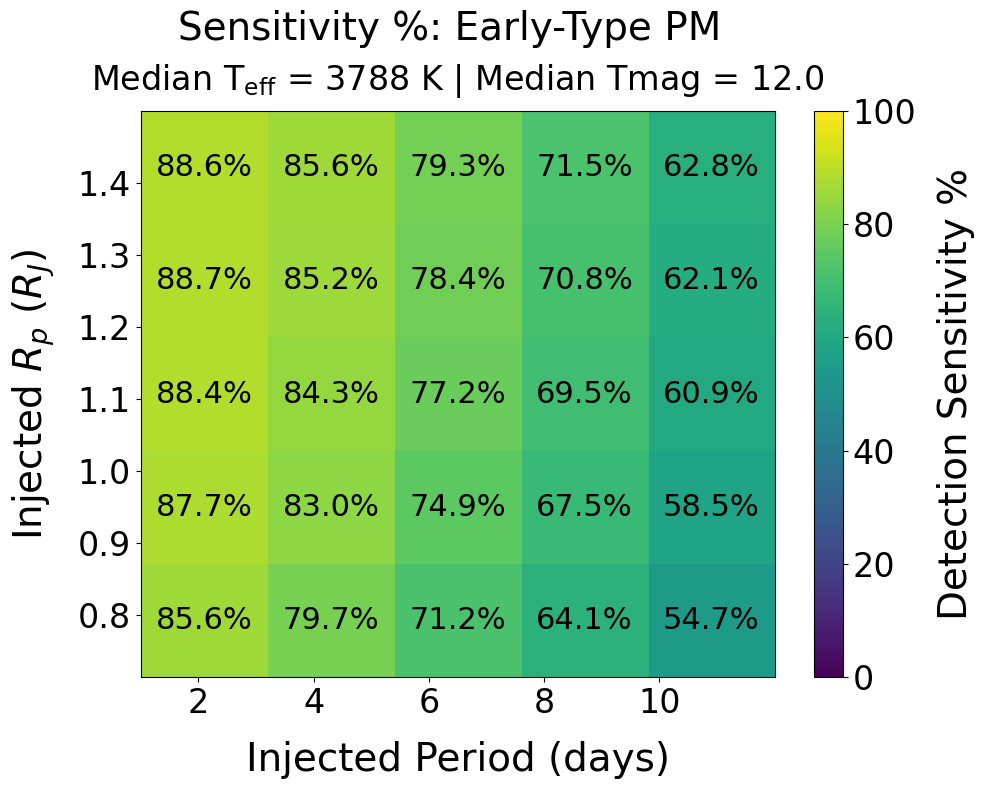}
    \end{subfigure}
    \hspace{0.01\textwidth}
    \begin{subfigure}[t]{0.30\textwidth}
        \centering
        \includegraphics[width=\textwidth]{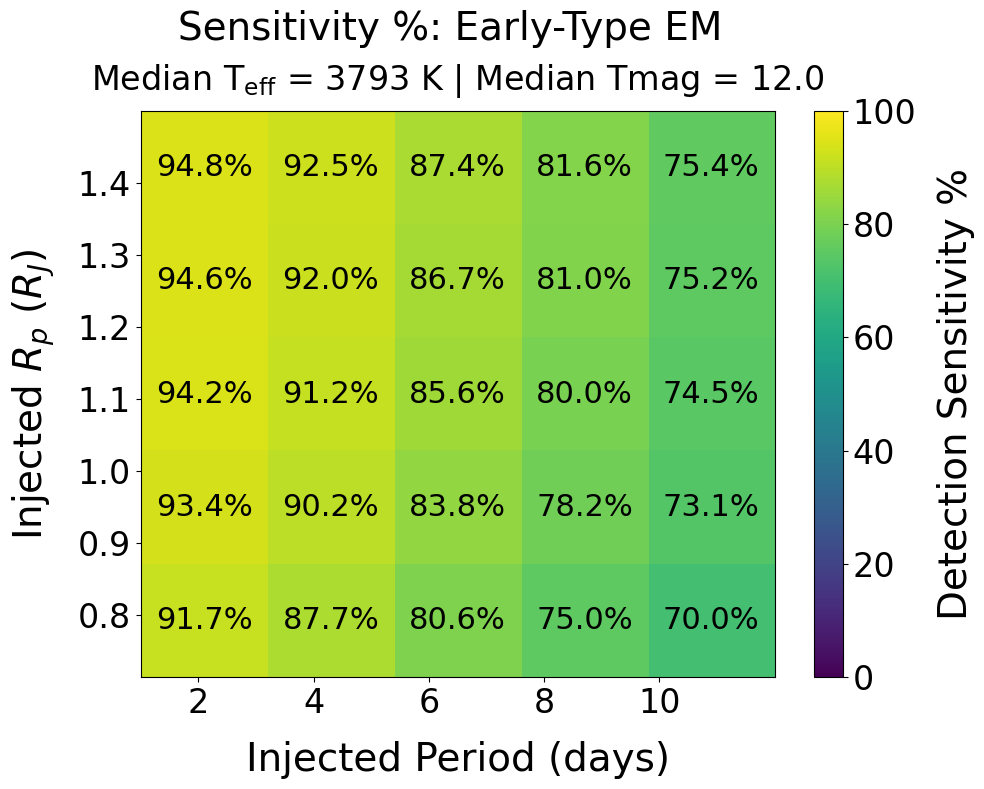}
    \end{subfigure}
    \hspace{0.01\textwidth}
    \begin{subfigure}[t]{0.30\textwidth}
        \centering
        \includegraphics[width=\textwidth]{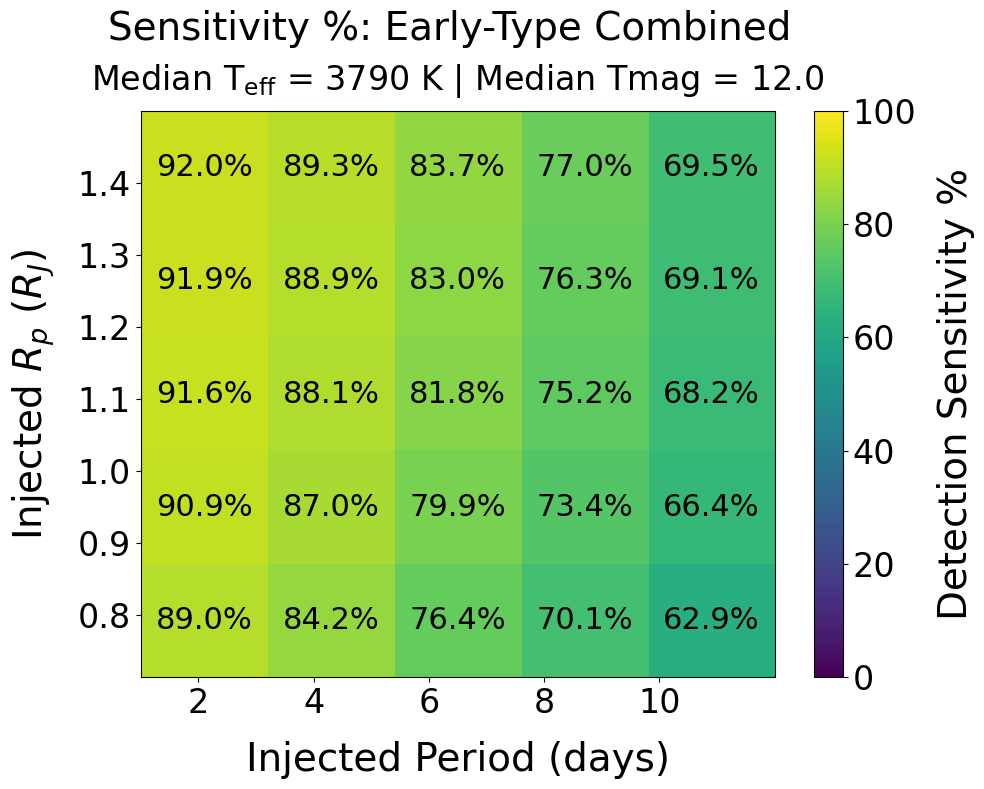}
    \end{subfigure}

    \vspace{1em}

    \begin{subfigure}[t]{0.30\textwidth}
        \centering
        \includegraphics[width=\textwidth]{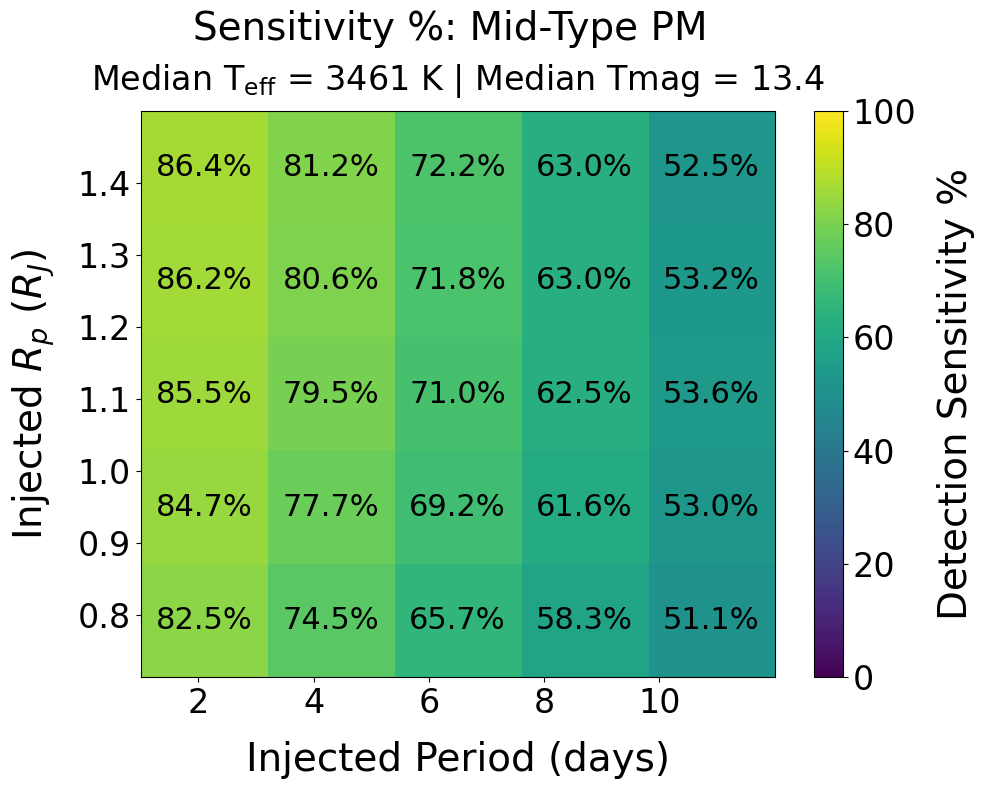}
    \end{subfigure}
    \hspace{0.01\textwidth}
    \begin{subfigure}[t]{0.30\textwidth}
        \centering
        \includegraphics[width=\textwidth]{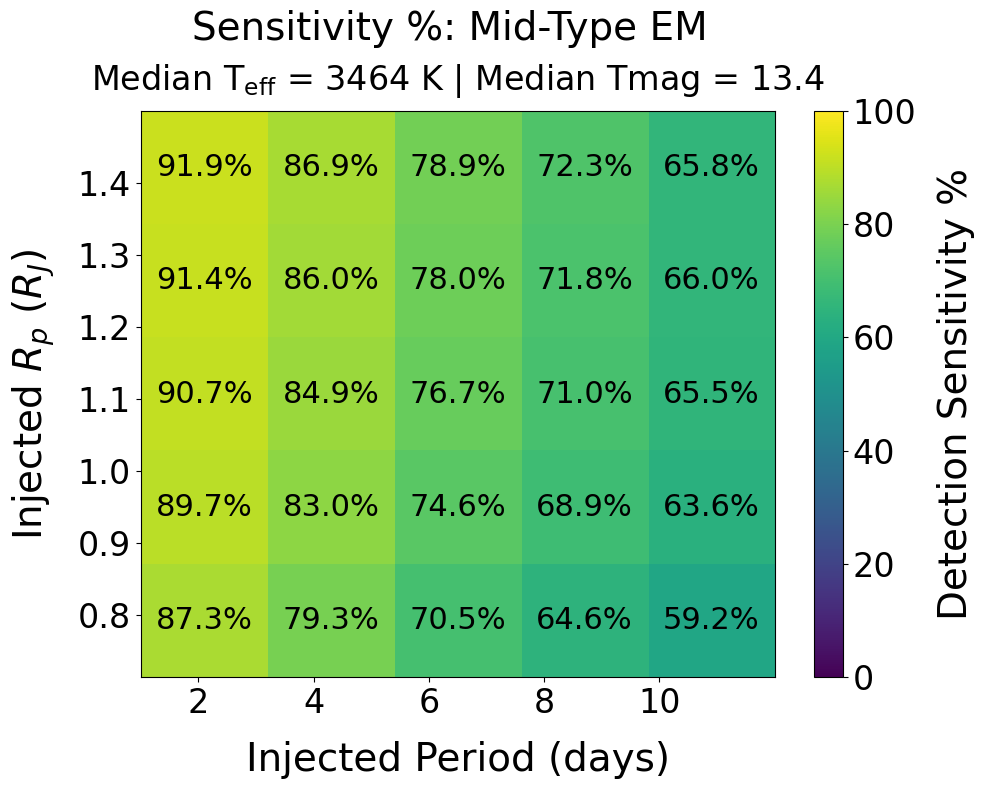}
    \end{subfigure}
    \hspace{0.01\textwidth}
    \begin{subfigure}[t]{0.30\textwidth}
        \centering
        \includegraphics[width=\textwidth]{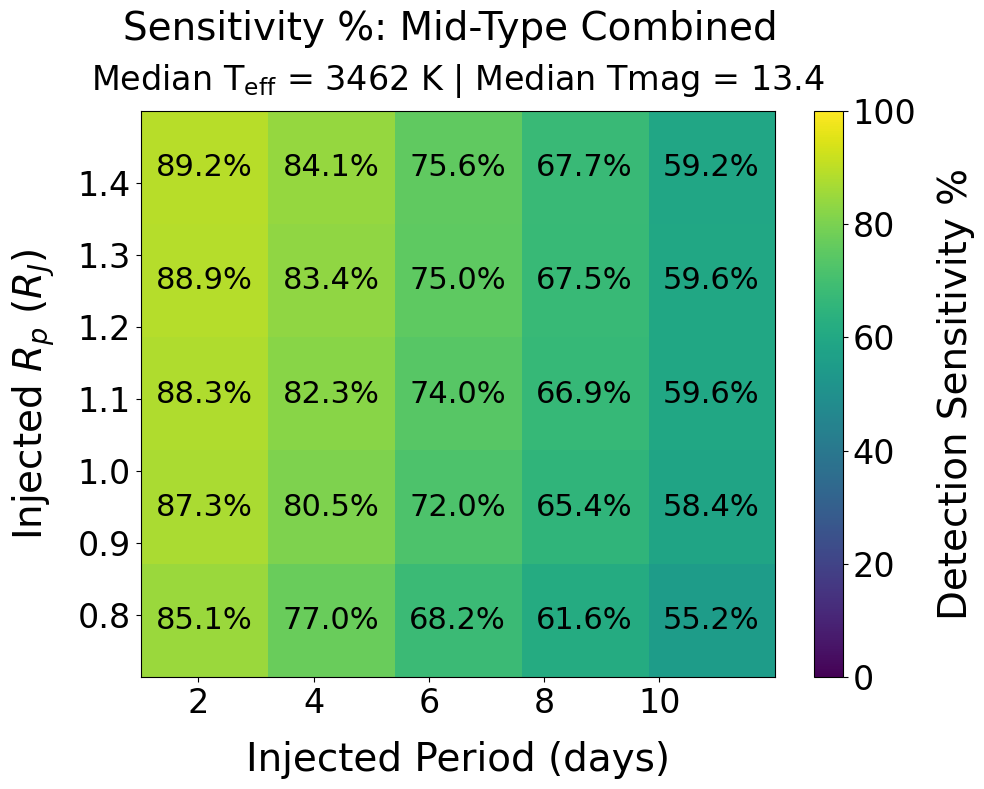}
    \end{subfigure}

    \vspace{1em}

    \begin{subfigure}[t]{0.30\textwidth}
        \centering
        \includegraphics[width=\textwidth]{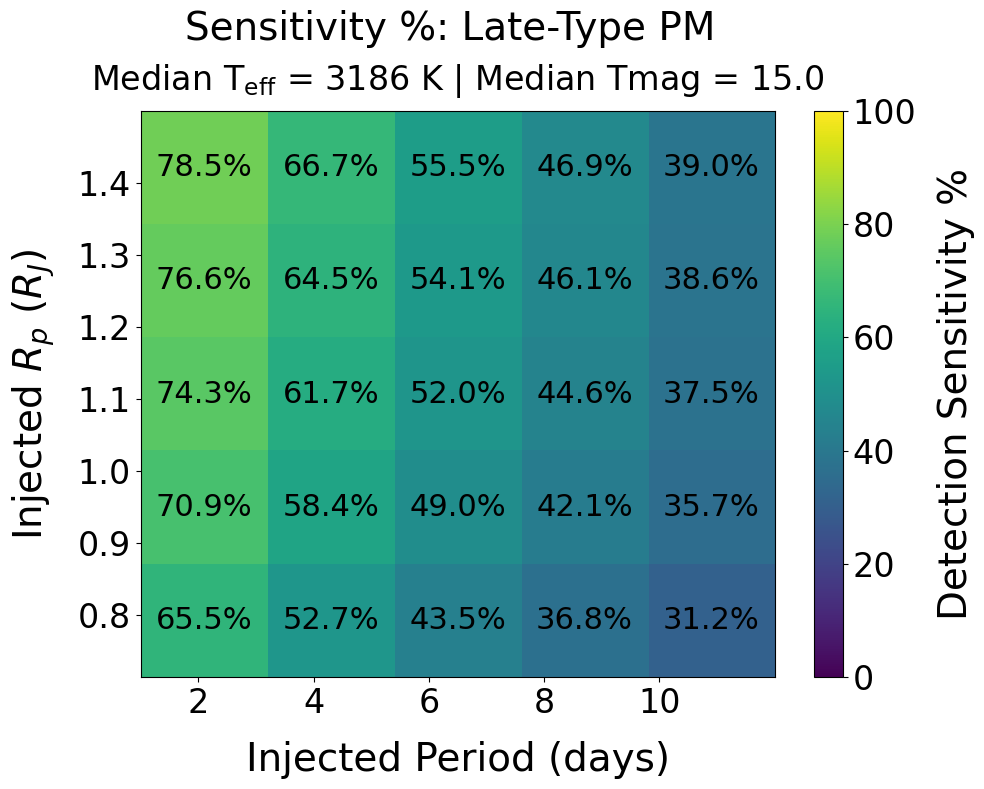}
    \end{subfigure}
    \hspace{0.01\textwidth}
    \begin{subfigure}[t]{0.30\textwidth}
        \centering
        \includegraphics[width=\textwidth]{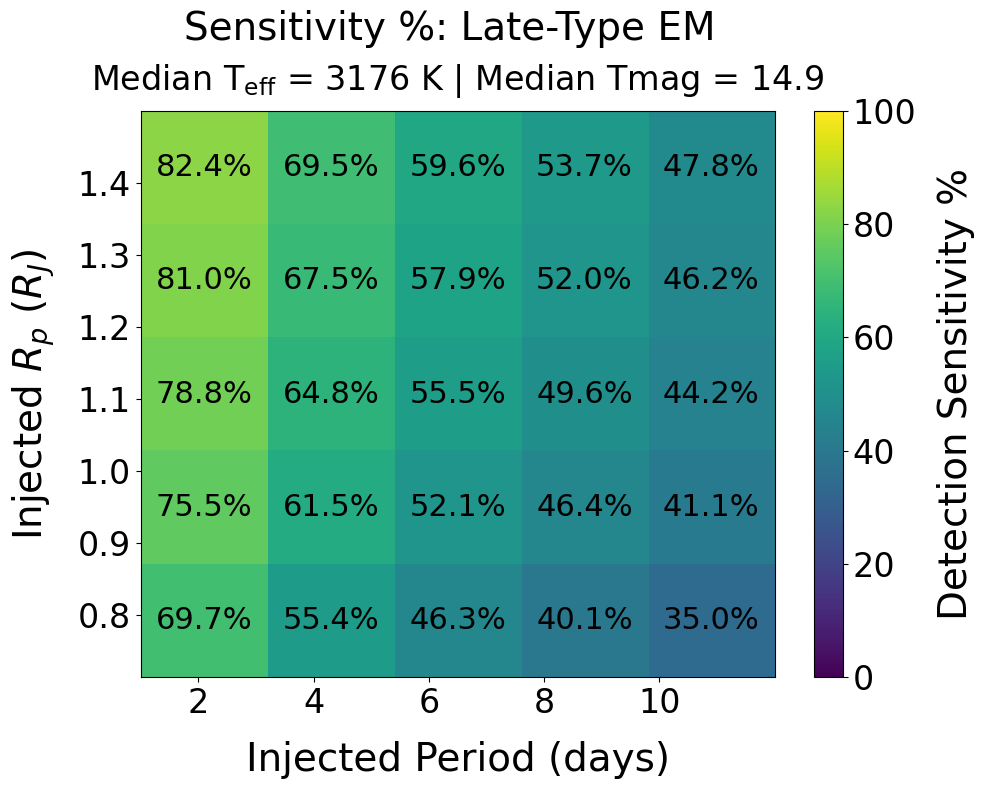}
    \end{subfigure}
    \hspace{0.01\textwidth}
    \begin{subfigure}[t]{0.30\textwidth}
        \centering
        \includegraphics[width=\textwidth]{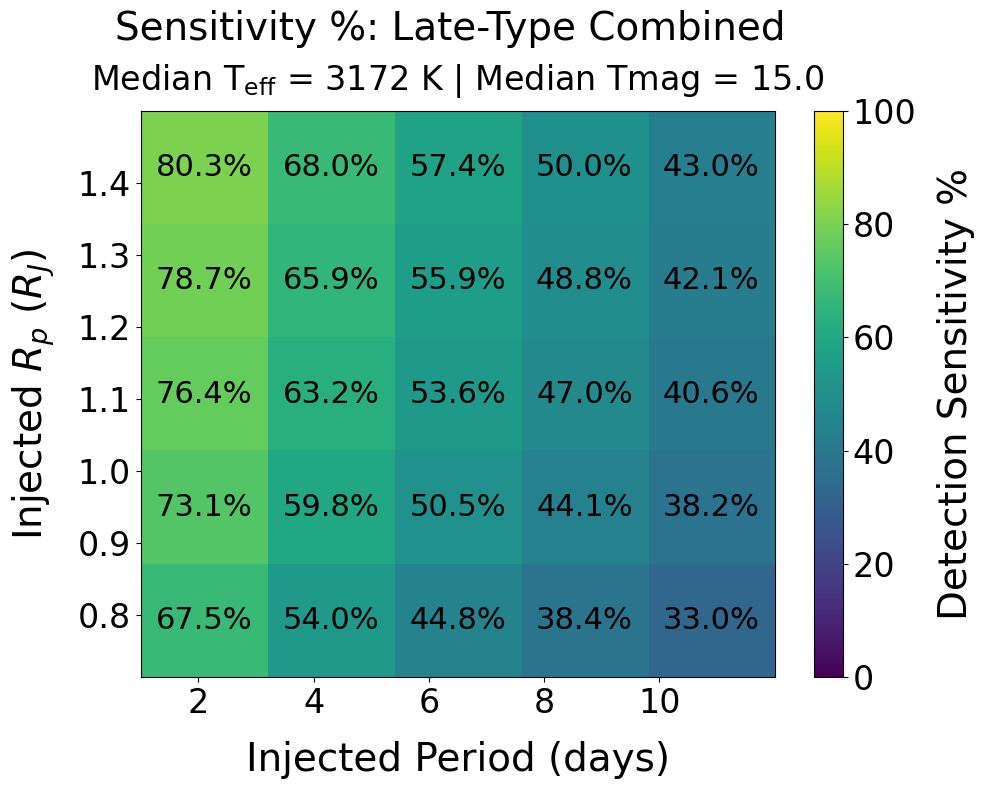}
    \end{subfigure}

    \caption{Detection sensitivity --- $S(P, R_p)$ --- grids for early, mid, and late-type M dwarfs (rows 1, 2, and 3). The primary (PM; Sectors 1-26) and extended mission 1 (EM1; Sectors 27-55) are shown in columns 1 and 2, whereas column 3 shows the combined average of the two. Each subfigure includes the median $T_{\text{eff}}$ and $T$-mag of the sampled stars.}
    \label{fig:sens_eml}
\end{figure*}

\begin{figure*}[ht]
    \centering

    \begin{subfigure}[t]{0.30\textwidth}
        \centering
        \includegraphics[width=\textwidth]{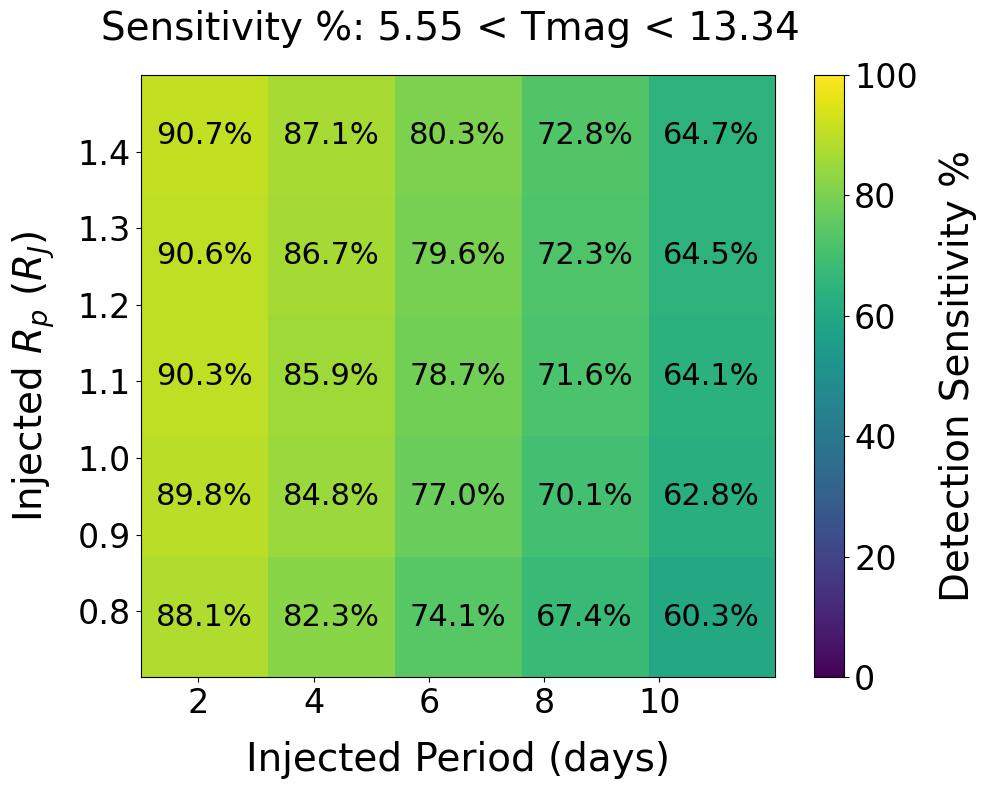}
    \end{subfigure}
    \hspace{0.01\textwidth}
    \begin{subfigure}[t]{0.30\textwidth}
        \centering
        \includegraphics[width=\textwidth]{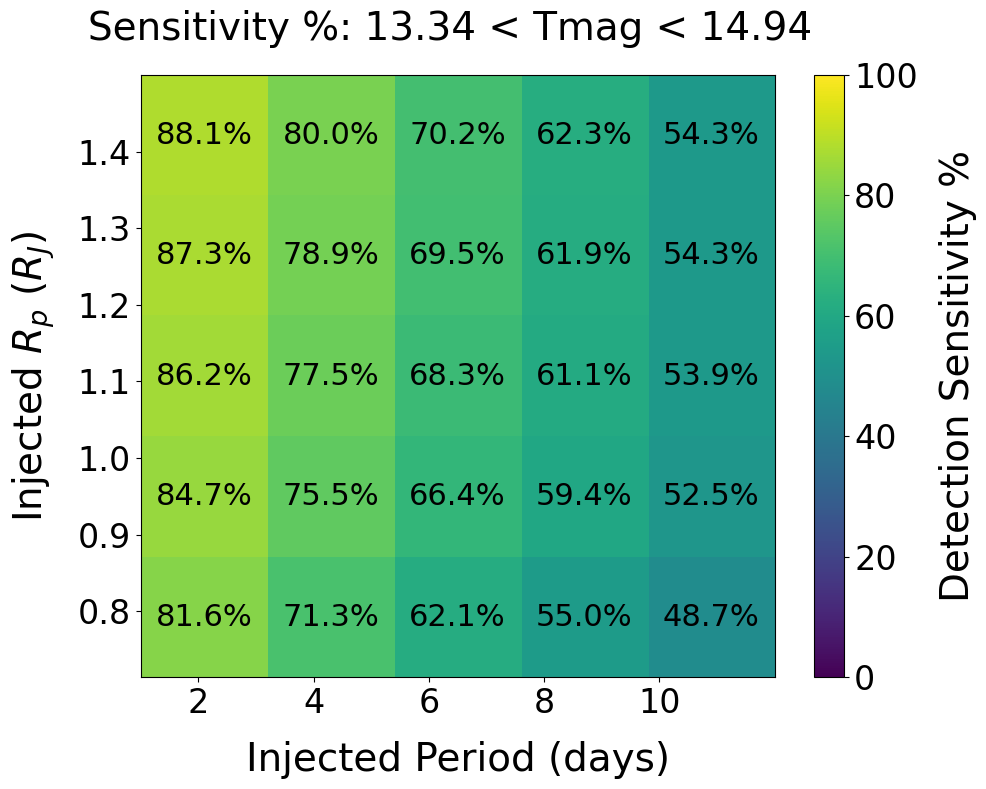}
    \end{subfigure}
    \hspace{0.01\textwidth}
    \begin{subfigure}[t]{0.30\textwidth}
        \centering
        \includegraphics[width=\textwidth]{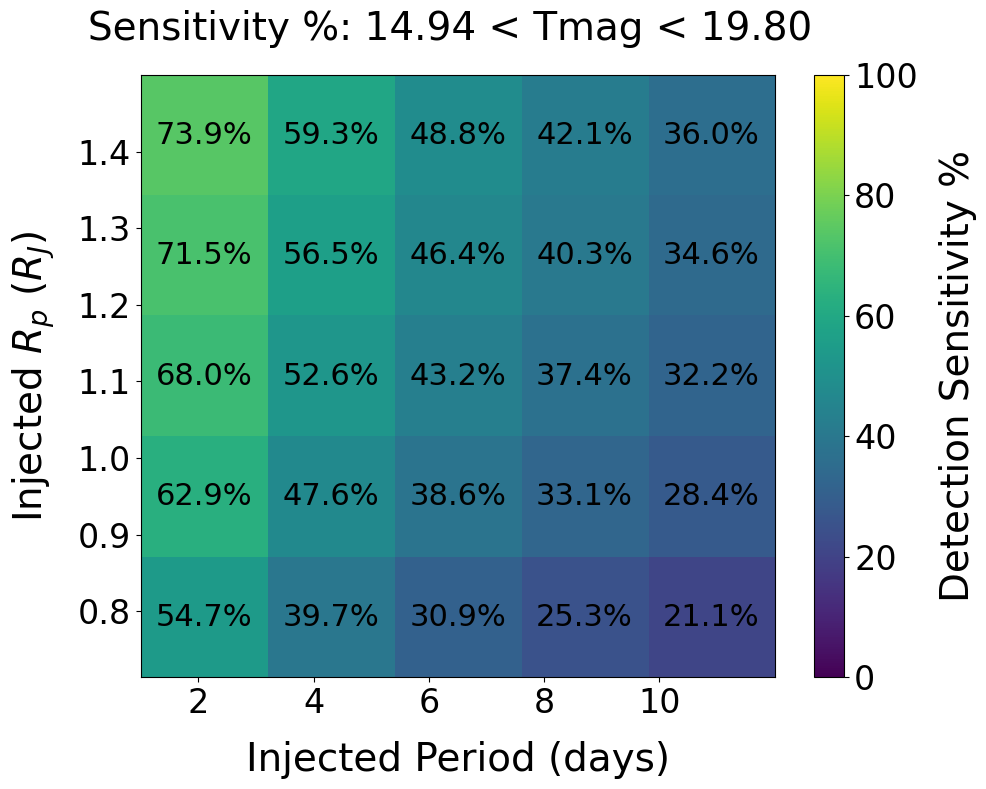}
    \end{subfigure}

    \caption{Detection sensitivity grids --- $S(P, R_p)$ ---  for equal thirds in $T$-mag. \textbf{Right:}  the sensitivity derived from hosts $14.94 < T\text{-mag} < 19.80$. \textbf{Center:} the sensitivity derived from hosts $13.34 < T\text{-mag} < 14.94$. \textbf{Left:}  the sensitivity derived from injected hosts $5.55 < T\text{-mag} < 13.34$. Each subfigure represents an equal third (17,009) of the total 51,027 sectors.}
    \label{fig:sens_tmag}
\end{figure*}

\begin{figure}[h]
    \centering
    \includegraphics[width=\linewidth]{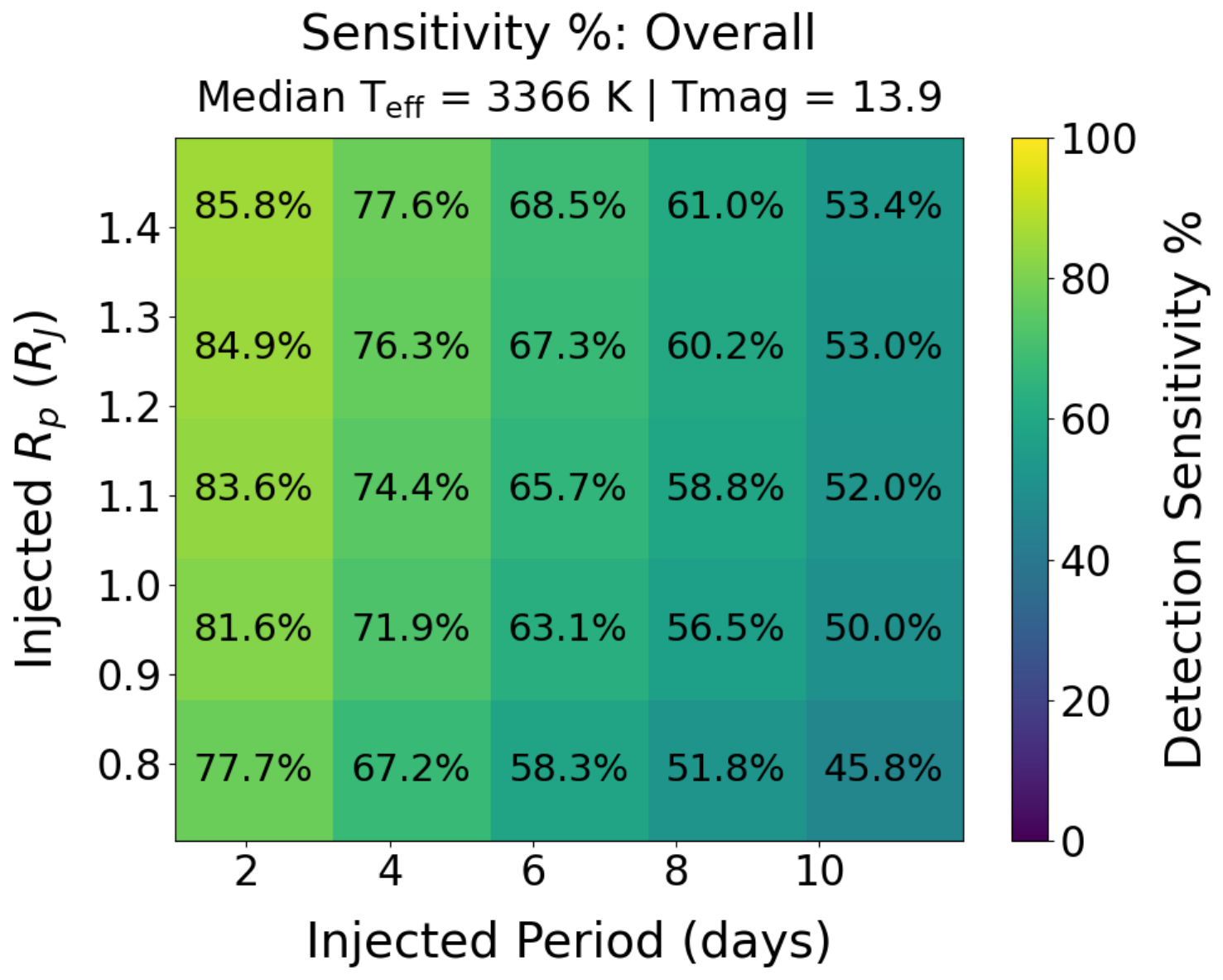}
    \caption{The detection sensitivity grid for our full sample of injected M-dwarfs, combined across spectral subtypes.}
    \label{fig:sens_all}
\end{figure}

\autoref{fig:sens_eml} shows our detection sensitivity --- $S(P, R_p)$ --- across the three M-dwarf spectral subtypes and the two TESS missions under consideration.  For a transit signal to be considered recovered, the corresponding sector must pass the following criteria: (a) the \texttt{TESS-miner} initial classification as a ``Good Fit" (see Section \ref{sec:flags}) should be false, (b) a BLS power threshold ($> 100$), (c) a minimum transit count ($N_\text{trans} > 1$), and (d) a radius constraint of $R_p < 2R_J$. We excluded the \texttt{OddEven} flag from this process to avoid computational overhead during runtime. The sensitivity grids were uniformly binned in orbital period and planet radius space.  Overall, our pipeline shows slightly higher sensitivity across all injected periods and $R_p / R_\star$ values in the EM1 data compared to the PM data, likely due to the higher cadence in EM1.

We also redistributed our injections across the top, middle and bottom third of \textit{T}-mag, visualizing their sensitivity grids in \autoref{fig:sens_tmag}. In this figure, we see that the sensitivity of our pipeline decreases with apparent brightness as expected. Meanwhile \autoref{fig:sens_all} shows the combined sensitivity across all spectral subtypes and missions.

\subsection{Completeness Grids}
\begin{figure*}[ht]
    \centering

    \begin{subfigure}[t]{0.3\textwidth}
        \centering
        \includegraphics[width=\textwidth]{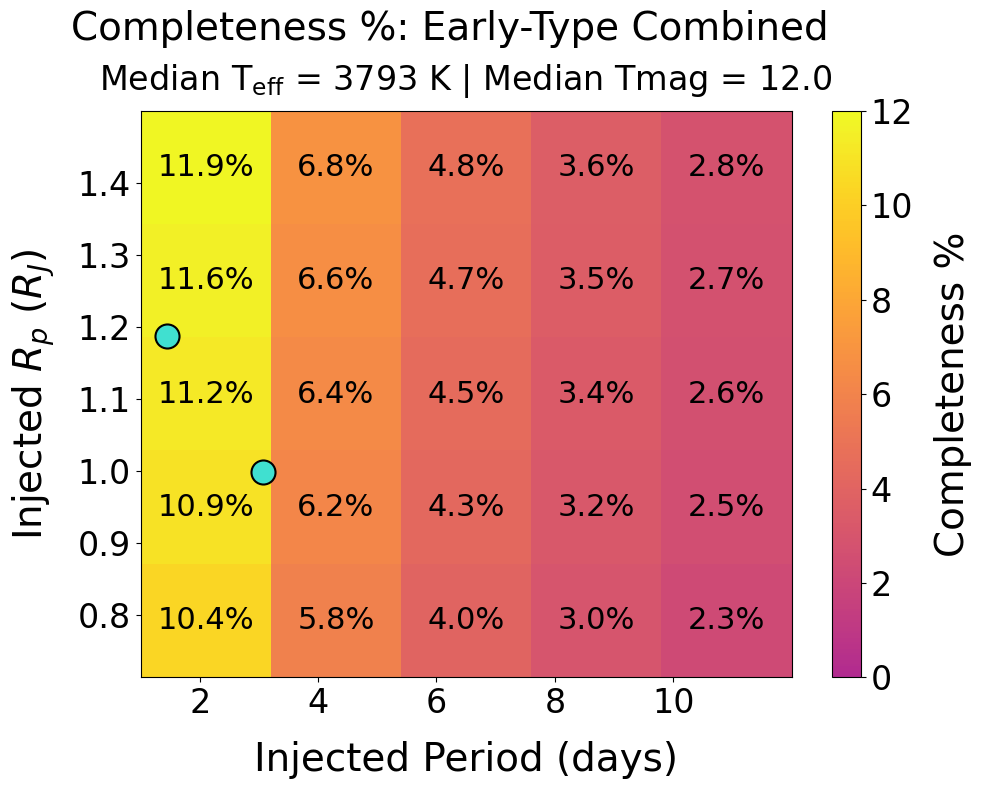}
    \end{subfigure}
    \hspace{0.01\textwidth}
    \begin{subfigure}[t]{0.3\textwidth}
        \centering
        \includegraphics[width=\textwidth]{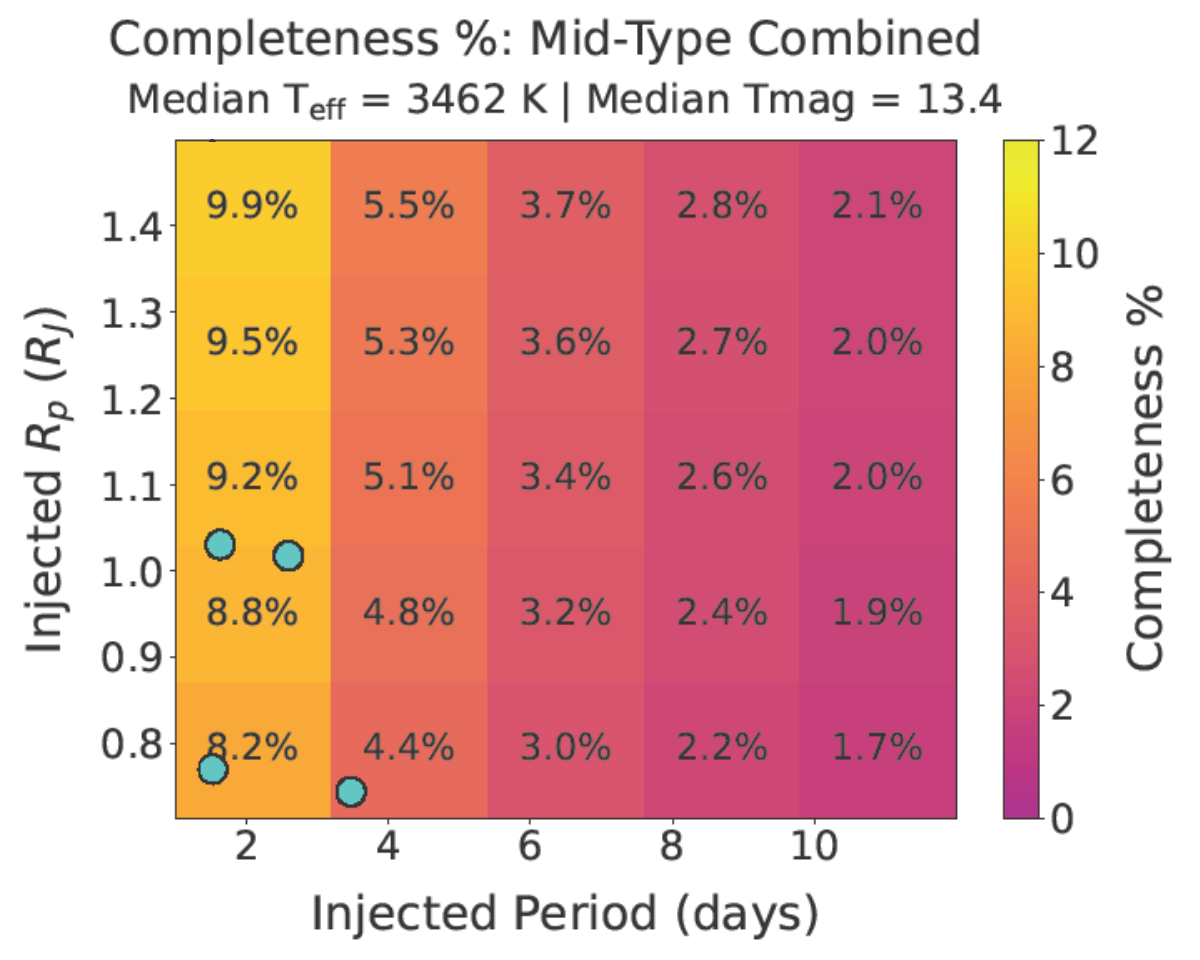}
    \end{subfigure}
    \hspace{0.01\textwidth}
    \begin{subfigure}[t]{0.3\textwidth}
        \centering
        \includegraphics[width=\textwidth]{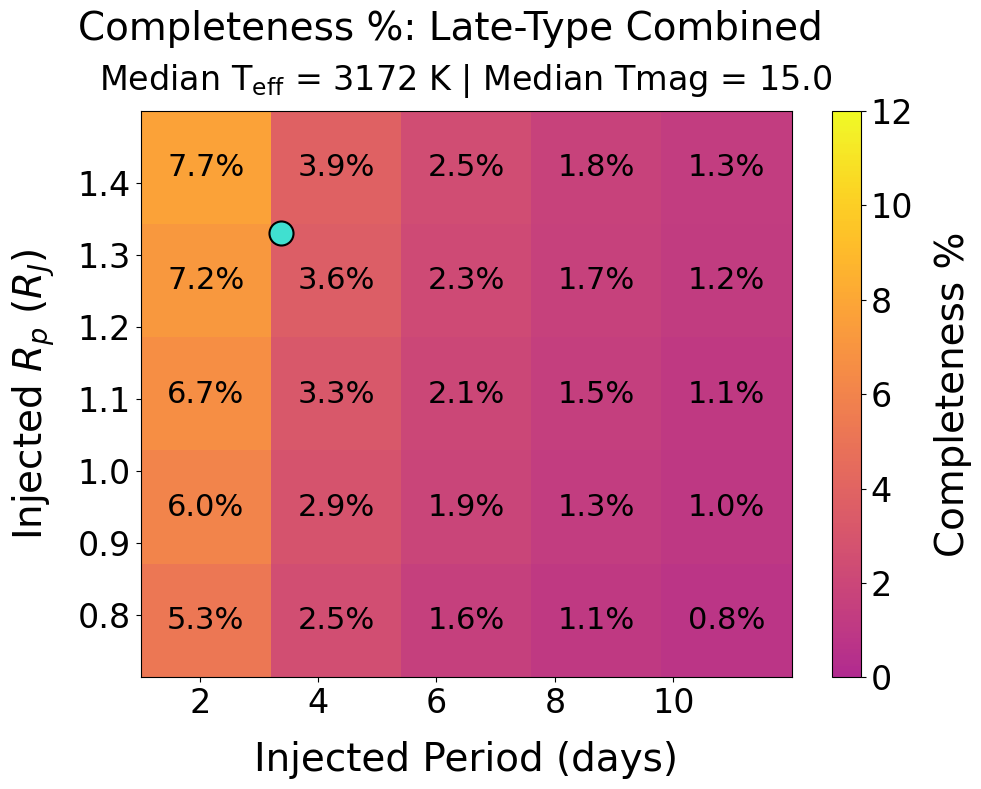}
    \end{subfigure}

    \caption{Completeness grids for early, mid, and late-type M stars observed in both the TESS primary and first extended missions (left to right). Each subfigure includes the median $T_{\text{eff}}$ and $T$ mag of the sampled stars. Confirmed planets are overplotted as turquoise circles. The new planet candidate is indicated with a white arrow.}
    \label{fig:comp_eml}
\end{figure*}

\begin{figure}[h]
    \centering
    \includegraphics[width=\linewidth]{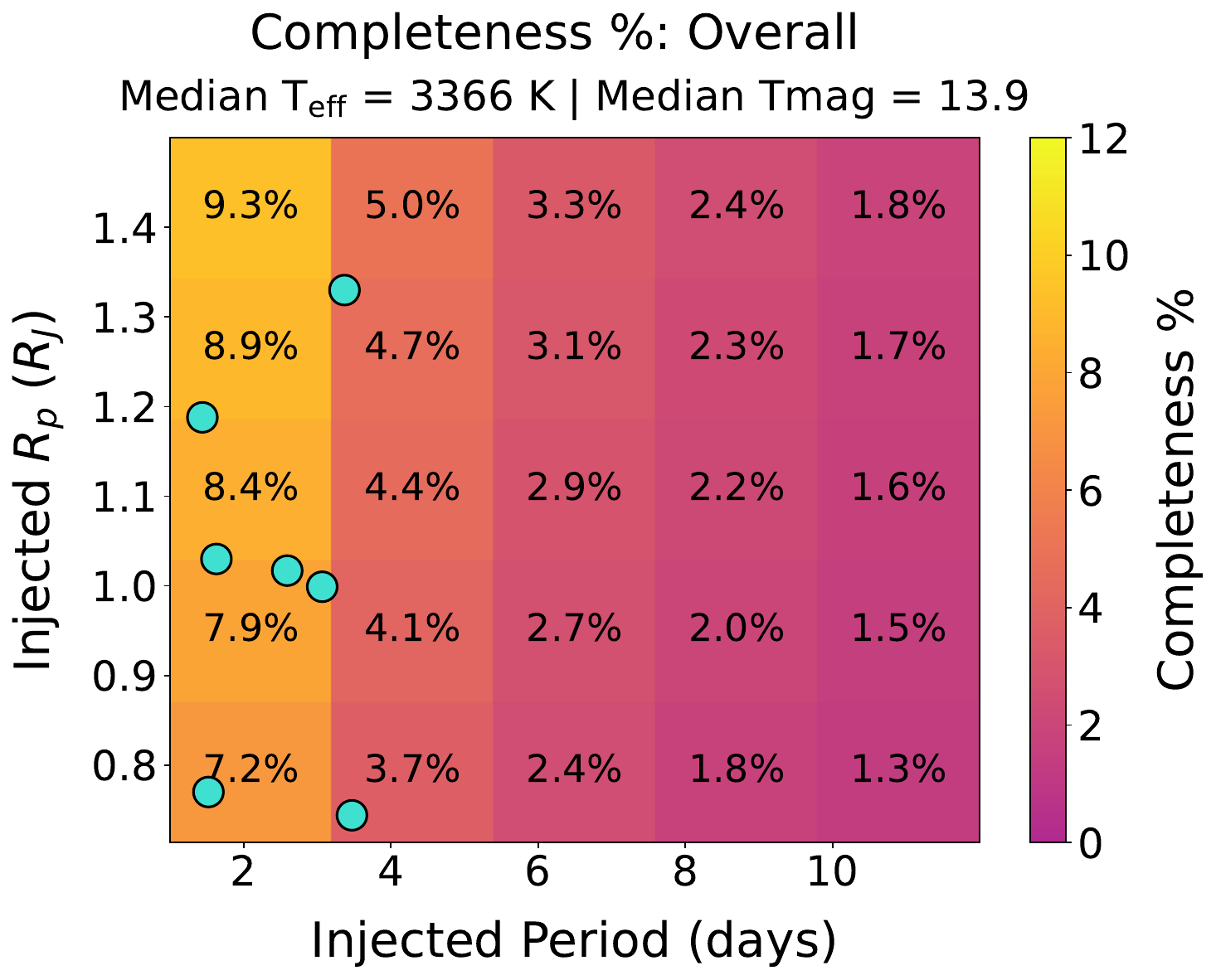}
    \caption{The completeness grid for our full sample of injected M-dwarfs, observed in both the TESS primary and first extended missions. Confirmed planets are indicated by blue circles.}
    \label{fig:comp_all}
\end{figure}

We calculated the geometric transit probability for each injected planet using:

\begin{align}\label{eq:transitprobability}
    \mathcal{P}_{tr} = \frac{R_\star + R_p}{a},
\end{align}

where $R_\star$ is the stellar radius; $R_p$ is the radius of the planet; and $a$ is the semi-major axis of the injected circular orbit, calculated using Kepler's third law. We then calculated the arithmetic mean of these geometric transit probabilities within each period–radius bin. These mean transit probabilities were multiplied by the corresponding recovery rates (i.e., the fraction of injections successfully recovered) to obtain the overall completeness, which is shown in \autoref{fig:comp_eml}. While the grids in this figure are binned for visualization purposes, we recalculated the completeness of each subsample without binning in order to avoid any biases in occurrence rates due to binning artifacts. For reference, the mean completeness percentages of the three spectral subtypes are $5.5\%$ (early), $4.2\%$ (mid), and $2.7\%$ (late). The overall mean completeness of our sample is $3.8\%$, and the completeness grid for all injected light curves (combined across spectral subtypes) is shown in \autoref{fig:comp_all}.


\subsection{Occurrence Rates} \label{sec:rates}

\subsubsection{Bayesian Method}

We adopt the simplified Bayesian modeling method suggested in Appendix B of \cite{hsu_improving_2018} to remain statistically robust despite the small planet sample. We place a Gamma distribution prior on the occurrence rate  $f_{occ}$ of each spectral subsample. The prior has a shape parameter $\alpha_0$ and a rate parameter $\beta_0$, such that the resulting posterior is the conjugate Gamma-Poisson solution for a Poisson process with mean $f_{occ}\cdot N_{*, \text{eff}}$. Because the number of planet detections is close to unity, the posterior for $f_{occ}$ is influenced by the choice of $\alpha_0$ and $\beta_0$. A more flexible alternative would be a hierarchical Bayesian modeling (HBM) approach, in which $f_{occ}$, $\alpha_0$, and $\beta_0$ vary as an assumed function of $T_{\text{eff}}$. However, we leave this exercise to future work with the full 200 pc M-dwarf sample. 

The mean of the Gamma posterior for $f_{occ}$ is: 

\begin{equation}\label{eq:bayesianmean}
    \mu_{f_{occ}} = \frac{\alpha_0 + n_{pl}}{\beta_0 + N_{*,\rm{TESS}}}
\end{equation}

and the uncertainty is 

\begin{equation}\label{eq:bayesiansigma}
    \sigma_{f_{occ}} = \sqrt{\frac{\mu_{f_{occ}}}{\beta_0 + N_{*,\rm{TESS}}}},
\end{equation}

where $n_{pl}$ is the number of planets in each sample, and $N_{*,\rm{TESS}}$ is the effective number of stars searched in our chosen period-radius range. We note that \autoref{eq:bayesiansigma} does not match Equation B5 from \cite{hsu_improving_2018}, since there is a typographical error in the latter,\footnote{The original equation found in \citet{hsu_improving_2018} confines the square root to the denominator.} which has been corrected here.

We did not adopt a Monte Carlo approach, such as that used by \citet{hsu_improving_2018}, to estimate the expectation value of the effective stellar sample size. Instead, we computed $N_{*,\text{TESS}}$ directly from our injection–recovery grid. For each spectral subsample, we calculated an effective completeness $C_{eff}$ as the arithmetic mean of the completeness values in those period–radius bins which host at least one planet  (see \autoref{fig:comp_eml} and \autoref{fig:comp_all}). We then defined
\begin{equation}
    N_{*,\text{TESS}}=C_{eff}\cdot N_{*,\text{raw}}
\end{equation}
where $N_{*,\text{raw}}$ is the number of stars in that subsample with usable TESS data. For example, for our full sample, we obtained $C_{eff}=7.16\%$, which resulted in $N_{*,\rm{TESS}}$ of $7.16\%~\times~145,612 = $ 10,425 stars. 

We adopt $\alpha_0$ = $\beta_0$ = 1, which corresponds to an exponential distribution prior on $f_{occ}$. The values of $f_{occ}$ corresponding to each spectral subsample are listed in the final column of \autoref{tab:occ_rates}. These are the values which we quote throughout the rest of this work. 

\subsubsection{Inverse Detection Efficiency Method (IDEM)}

For comparison, we also estimated the occurrence rate within each spectral subsample using the Inverse Detection Efficiency Method (IDEM) described by \cite{howard_planet_2012}, and subsequently \cite{2015ApJ...810...95C} as well as others. Given the sparsity of planets, we did not estimate occurrences in each radius-period cell separately. Here, for the $n_{pl,j}$ planets in spectral subtype~\textit{j}, each planet $i$ contributes to the occurrence $f_{occ,j}$ as follows:

\begin{equation} \label{eq:focc}
\begin{split}
    f_{occ,j} = \sum_i^{n_{pl,j}} \frac{1}{\mathcal{P}_{tr, i} \cdot S_j(P, R_p) \cdot N_{\star,j}} \cdot FPP_i, \\ \forall j \in \{\textrm{Early, Mid, Late}\}
\end{split}
\end{equation}

where $S_j(P, R_p)$ is the detection sensitivity in a period-radius bin as shown in \autoref{fig:sens_eml}, and $N_{\star,j}$ are the number of stars, in a given spectral subtype $j$. $\mathcal{P}_{tr, i}$ refers to the transit probability for planet $i$ from \autoref{eq:transitprobability}. 

We accounted for errors in the period and radius estimates of all confirmed planets by bootstrapping the above equation across 5,000 iterations and accounting for the bootstrap-induced uncertainty ($\sigma_{\text{boot}}$). Our seven confirmed planets have a false positive probability (FPP) of 0. We note that bootstrap iterations in which a planet fell outside our injection bounds did not contribute to the occurrence rate.

For the uncertainty in $f_{occ,j}$, we follow recommendations from Appendix A by \cite{hsu_improving_2018}, added in quadrature with errors introduced by bootstrapping:

\begin{equation}
    \sigma_{f_{occ,j}} = \sqrt{\Big(\frac{f_{occ,j}}{\sqrt{n_{pl,j}}}\Big)^2 + \sigma_{\text{boot}}^2}
\end{equation}

\begin{deluxetable*}{c|c|c|c|c|c|c|}
\setlength{\tabcolsep}{7pt}  
\tablecaption{Occurrence rates of GEMS by spectral type\label{tab:occ_rates}}
\tablewidth{\linewidth}          
\tabletypesize{\normalsize}
\tablehead{
\colhead{\textbf{Type}} &
\colhead{\textbf{Median} $\mathbf{M_\star ~(M_\odot)}$} &
\colhead{$\mathbf{N_\star}$} &
\colhead{$\mathbf{N_{\star \text{, eff}}}$} &
\colhead{$\mathbf{N_p}$} &
\colhead{$\mathbf{IDEM}$ $\mathbf{f_{\rm occ}}$} &
\colhead{$\mathbf{\Gamma(1,1)}$* $\mathbf{f_{\rm occ}}$} [\%]}

\startdata
\hline
Early & 0.54 & 27,995 & 3,149 & 2 & $0.067 \pm 0.047$ & $0.118 \pm 0.068$\\
Mid   & 0.35 & 37,948 & 2,903 & 4 & $0.139 \pm 0.069$ & $0.153 \pm 0.069$ \\
Late  & 0.18 & 79,669 & 2,868 & 1 &           $0.032 \pm 0.032$ & $0.036 \pm 0.024$ \\ 
\hline
Full  & 0.25 & 145,612 & 9,902 & 7 &           $0.065^{+0.025}_{-0.027}$ & $0.068 \pm 0.024$ \\
\hline
\enddata
\tablecomments{Each row lists the number of stars analyzed ($N_\star$), the effective number of stars ($N_{\star\text{, eff}}$), and the number of planets ($N_p$) per mass bin. Occurrence rates are computed with a simplified Bayesian approach with a $\Gamma(\alpha_0,\beta_0)$ prior (see Section~\ref{sec:rates}) and, for comparison, with the Inverse Detection Efficiency Method (IDEM). Stars lacking usable TGLC data are excluded from $N_\star$. IDEM values are not adopted as the final results in this work, but are presented as a comparison to the hierarchical Bayesian method.}
\tablecomments{* - Adopted}
\end{deluxetable*}

\begin{figure*}
    \centering
    \includegraphics[width=0.8\textwidth]{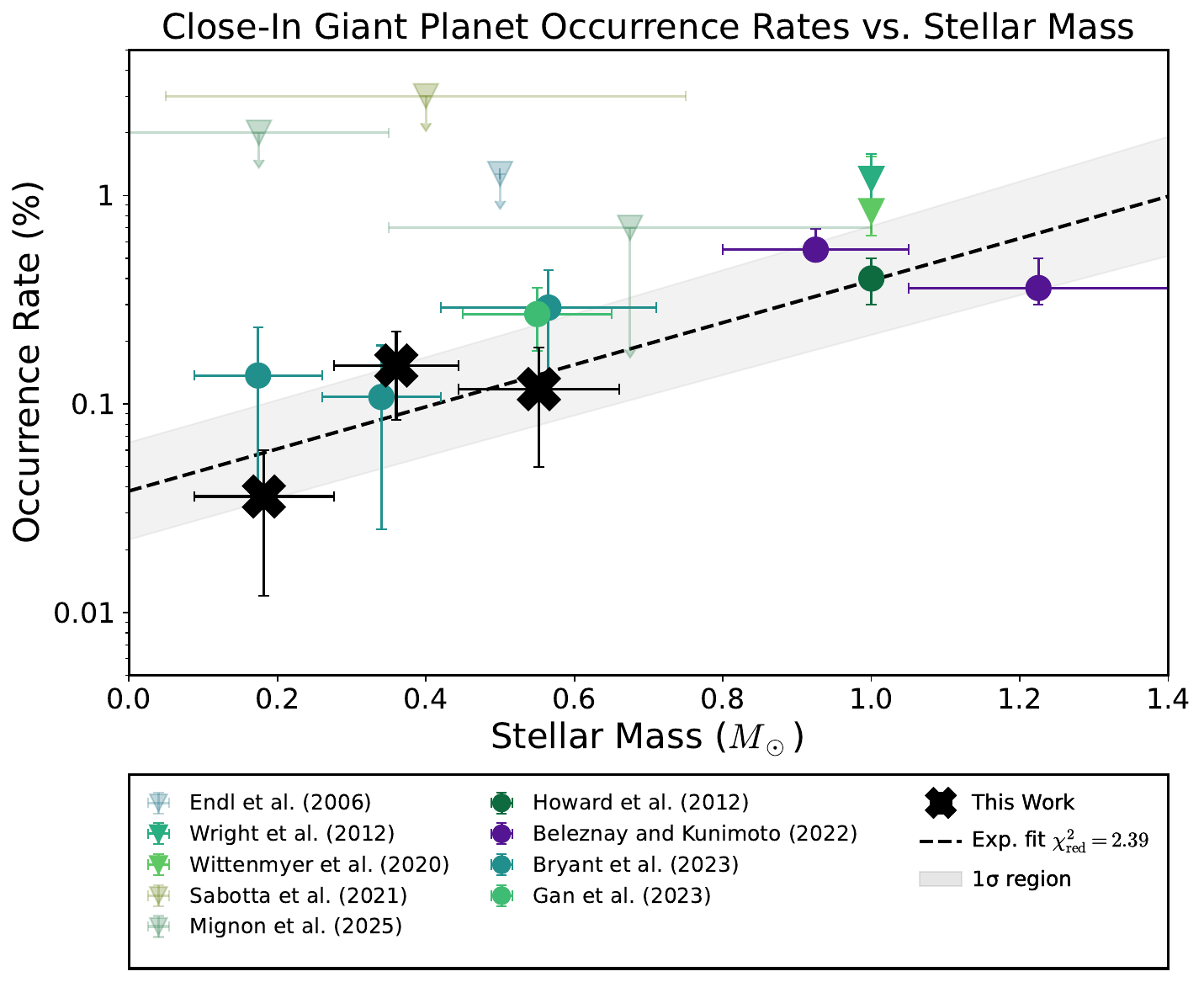}
    \caption{The occurrence rates of close-in giant planets around FGKM dwarfs based on RV searches (triangles),\footnote{The inclusion of RV-based occurrence rates in this figure served purely as a qualitative comparison. We leave quantitative comparisons of RV- and transit-based occurrence rates to future work.} transit searches (circles), and this work (filled crosses). Upper limits are denoted by downward arrows and slight transparency, as in the case of \citet{endl_exploring_2006}, \citet{howard_planet_2012}, and \citet{mignon2025}. The upper limit from \citet{endl_exploring_2006} is shown at a representative stellar mass of 0.5 \solmass{}, whereas those for \citet{wright_frequency_2012}, \citet{howard_planet_2012}, and \citet{wittenmyer_cool_2020}  are shown at 1 \solmass{} given that the latter consist of a sample of FGK dwarfs.  Finally, the dashed black line and associated grey error envelope represent an exponential fit to the black filled crosses from this work and the purple points from \citet{beleznay_exploring_2022}. Fitting to these works allows us to span the stellar mass axis using only transit photometry based studies in TESS data within a similar period and radius range to our own work.}
    \label{fig:occrates_lit}
\end{figure*}

We note here that the contribution to the occurrence rate error from bootstrapping made up $<1\%$ of the total error for early- and late-type bins, due to the small number of planets occurring on the edge of a given completeness bin (see \autoref{tab:occ_rates}). Values of $f_{occ}$ calculated using this method are listed in \autoref{tab:occ_rates}. 

While IDEM is computationally and intuitively simple, it has limitations, especially in the marginal detection regime where the transit S/N is comparable to the detection threshold, (i.e., for small planets, as has been highlighted by \citet{hsu_improving_2018}). However, for large planets such as those under consideration here, the transit S/N is typically much larger than the detection threshold, as evidenced by our radius uncertainties in \autoref{tab:finalsample}. \textbf{Here, the biases induced by IDEM compared to other occurrence rate (and uncertainty) estimation techniques such as Approximate Bayesian Computing \citep[ABC; see Figure 6 in][]{hsu_improving_2018} are expected to be smaller.}

\section{Discussion} \label{sec:discussion}

\subsection{Contextualizing the occurrence of GEMS}

\begin{figure*}[t]
    \centering
    \begin{subfigure}[t]{0.48\textwidth}
        \centering
        \includegraphics[width=\linewidth]{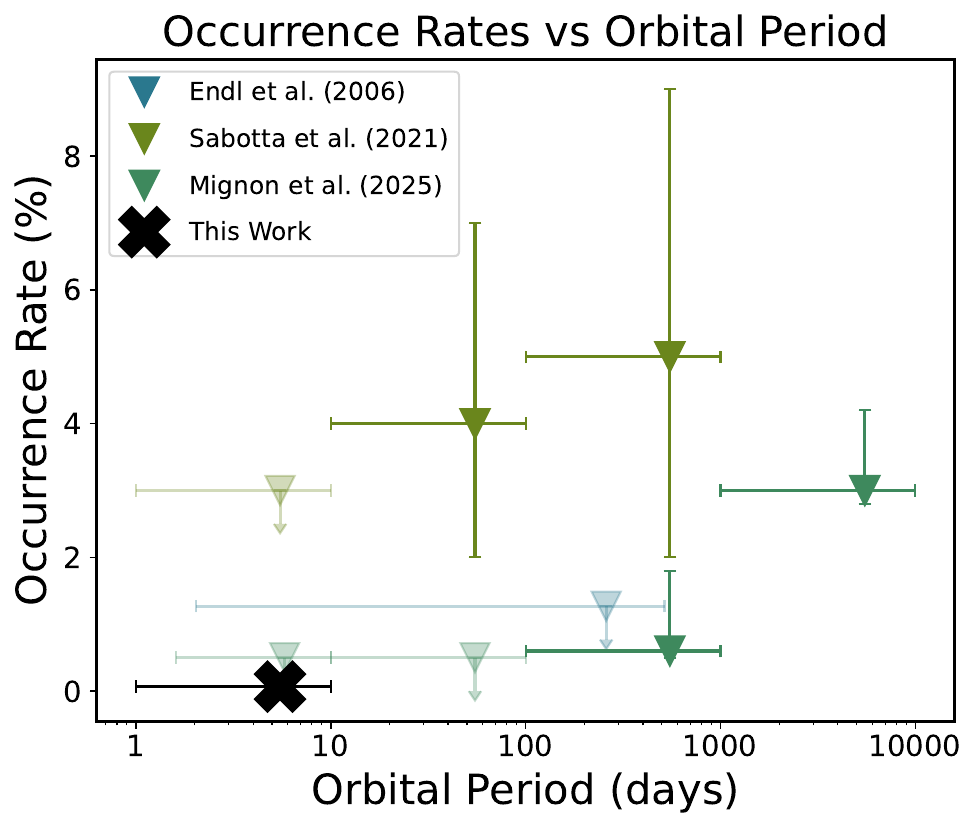}
        \caption{Occurrence rates of GEMS across orbital period}
        \label{fig:occ_per}
    \end{subfigure}
    \hfill
    \begin{subfigure}[t]{0.48\textwidth}
        \centering
        \includegraphics[width=\linewidth]{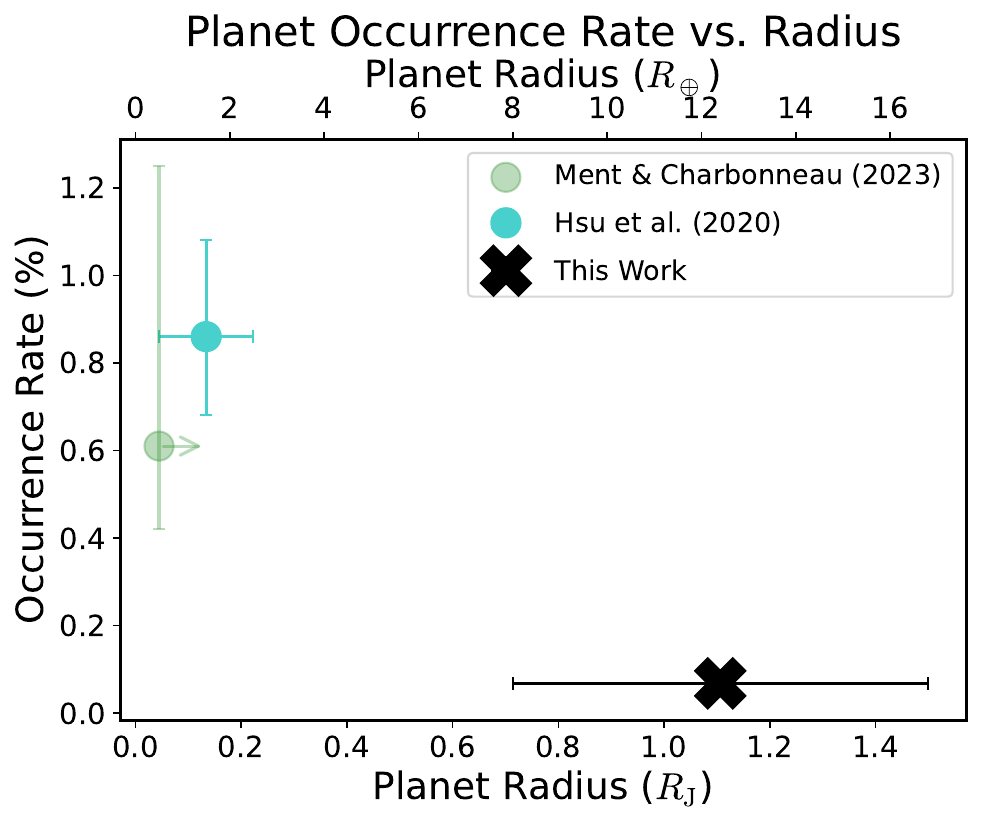}
        \caption{Occurrence rates of planets orbiting M-dwarfs across planet radius}
        \label{fig:occ_rad}
    \end{subfigure}
    \caption{Following the convention from \autoref{fig:occrates_lit}, we show \textbf{a)} occurrences of giant planets around M-dwarfs likely increases monotonically with orbital period, with close-in hot Jupiters being the rarest of all. While the total search range associated with this work is $1.0 < P < 10.0$ days, our occurrence rates were calculated using planets orbiting with $1.0 < P < 5.0$ days. We choose to use the former range for this illustration. \textbf{b)} Occurrences as a function of planet radius. This figure suggests that the occurrence rate of planets orbiting M-dwarfs decreases with planet radius. No transit-based occurrence rates are available for planets with radii $4 ~\mathrm{R_\oplus} < R_p < 8 ~\mathrm{R_\oplus}$ (Neptunes and sub-Saturns) around M-dwarfs. Limits are denoted by transparency and arrows,\footnote{\citet{Ment2023} reported occurrence rates for planets with periods between 0.5 and 7.0 days, slightly different from our nominal range of 1.0–10.0 days. However, because our rates are calculated for planets with periods between 1.0 and 5.0 days, the two remain directly comparable. Similarly, \citet{hsu_occurrence_2020} presented rates for periods between 0.5 and 8.0 days.} and RV-based occurrence rates are denoted by triangles. In both subplots, the nominal occurrence rate found in this work is indistinguishable from the maximal value.}
    \label{fig:combined_occ}
\end{figure*}

\autoref{fig:occrates_lit} shows our occurrence rates plotted against stellar mass alongside rates from previous works, spanning FGKM-dwarfs \citep{endl_exploring_2006, howard_planet_2012, wright_frequency_2012, wittenmyer_cool_2020, sabotta_carmenes_2021, beleznay_exploring_2022, mignon2025} as well as previous TESS GEMS occurrence rates from \citep{bryant_occurrence_2023, gan_occurrence_2023}. It is clear from this figure that the occurrence rates of giant planets across FGKM-dwarfs are consistent with a positive, monotonic relationship to stellar mass. We note that we have chosen the occurrence rate of $0.4\%\pm0.1\%$ from \citet{howard_planet_2012}, which is specific to planets between $8$ and $32 \, R_\oplus$ with $P<10$ days, as it mimics our search bounds the closest. 


To compare how hot-Jupiter occurrence varies with stellar mass across different surveys, we fit the exponential relation $f_{\mathrm{occ}}= a\cdot 10^{b \cdot M_\star}$ to (i) the nominal occurrence rates from this work, (ii) the occurrence rates reported by \citet{bryant_occurrence_2023}, and (iii) those of \citet{gan_occurrence_2023}. All three data sets were supplemented by the occurrence rates provided by \citet{beleznay_exploring_2022} for transiting Jupiters around FGK dwarfs. We assumed a Gamma prior for the occurrence rates derived in this work and fitted Beta priors to those from the other three studies. We then drew $10,000$ pseudo-rates from each prior distribution and fitted the exponential parameters $a$ and $b$ for each iteration. The median values of the distributions of $a$ and $b$, their $1\sigma$ uncertainties, and the reduced \(\chi^2\) value of the resultant fit are listed in \autoref{tab:expfits}. The exponential fit corresponding to this work is included, along with its $1\sigma$ region, in \autoref{fig:occrates_lit}. 

Our fit exhibits the steepest slope of the three, strengthening the evidence that the frequency of close-in giant planets is much smaller around low-mass stars than around their higher-mass counterparts.  Such a pronounced mass dependence aligns with the predictions of core-accretion theory, which suggests an ease of forming giant-planet cores around more massive protoplanetary discs (and hence higher-mass host stars).

The two leading theories of giant planet formation are core accretion \citep{Mizuno1980, bodenheimer_calculations_1986, Pollack96, ikoma_formation_2000} and gravitational instability \citep{boss_giant_1997, boss_forming_2023}. In the core accretion model, material from the protoplanetary disk accumulates onto a planetesimal, eventually forming a gas giant with a solid core. In contrast, the gravitational instability model posits that portions of the disk undergo rapid collapse under their own gravity, forming gas giants without rocky cores. Both scenarios require sufficiently massive disks, making giant planet formation challenging around low-mass stars.

While the observed increase in giant planet occurrence with stellar mass alone does not decisively favor one theory, recent findings show that GEMS preferentially orbit metal-rich M-dwarfs \citep{gan_metallicity_2024, han_toi-5344_2024}. This mirrors the planet–metallicity correlation established for FGK stars \citep{fischer_planet-metallicity_2005, maldonado_hades_2020, Osborn2020}, suggesting a shared formation pathway. Metal-rich stars host more solids necessary for core growth \citep{ida_toward_2004_2}, strengthening the case for core accretion. Additionally, \cite{kanodia_transiting_2024} show a commonality in the bulk properties of transiting giant planets around FGKM dwarfs, which also hints at similar formation pathways. Together, these provide tentative support for core accretion over gravitational instability as the dominant formation pathway GEMS.

\begin{deluxetable}{p{3cm}|c|c|c|c|c}
\setlength{\tabcolsep}{6pt} 
\tabletypesize{\normalsize}
\tablewidth{\linewidth}  
\tablecaption{Coefficients for the exponential fit $f_{\text{occ}}=a\cdot10^{b\cdot M_*}$ for the hot-Jupiter occurrence rate as a function of stellar mass.\label{tab:expfits}}
\tablehead{
\colhead{\centering Ref.} &
\colhead{$a$} &
\colhead{$\sigma(a)$} &
\colhead{$b$} &
\colhead{$\sigma(b)$} &
\colhead{$\chi^2_\text{red}$}
}
\startdata
This work                          & 0.038 & 0.021 & 1.009 & 0.273 & 2.39 \\
\citet{bryant_occurrence_2023}     & 0.090 & 0.059 & 0.707 & 0.324 & 2.46 \\
\citet{gan_occurrence_2023}        & 0.265 & 0.166 & 0.250 & 0.288 & 8.46 \\
\enddata
\tablecomments{The steeper slope in our TESS‐based fit indicates a stronger decline in giant-planet frequency toward low-mass hosts, consistent with core-accretion expectations.}
\end{deluxetable}

\autoref{fig:occ_per} shows the occurrence rates of GEMS from previous studies and from this work, plotted against orbital period \citep{endl_exploring_2006, sabotta_carmenes_2021, mignon2025}. The figure suggests a possible positive correlation between period and occurrence rate, though caution is warranted when combining results from different detection techniques. Still, such comparisons can be informative: \citet{fernandes_hints_2019} reported a similar rise in occurrence with orbital period for FGK giant planets, with a turnover near the ice line. Identifying a comparable turnover for GEMS would be particularly interesting. While current estimates for long period GEMS remain uncertain due to smaller RV sample sizes, Gaia DR4 astrometric detections may help by probing intermediate separations \citep{perryman_astrometric_2014, 2025AJ....169..107S}. Furthermore, observations by the Nancy Grace Roman Space Telescope \citep[Roman;][]{Penny2019} have the potential to detect transiting warm and cold Jupiters \citep{Wilson_2023}.

Finally, \autoref{fig:occ_rad} displays the occurrence rates of planets orbiting M-dwarfs across planet radius. There is a considerable lack of data in the super-Neptune radius range, highlighting the necessity for further research in that regime. However, our result for planets $0.7 ~\mathrm{R_J} \leq R_p \leq 1.5 ~\mathrm{R_J}$ is significantly smaller than planet occurrence rates at smaller radii.

\subsection{Comparison with previous works}

This study differs from previous work from \cite{gan_occurrence_2023} and \cite{bryant_occurrence_2023} in several key ways. First, the sample of M-dwarfs analyzed in this paper is the largest and includes a more comprehensive range of low-mass M-dwarfs ($M_\star/M_\odot < 0.4$) compared to previous transit-based GEMS searches. We achieved this by using \texttt{tglc}-produced light curves, which include fainter targets than the QLP or TESS-SPOC pipelines (\autoref{tab:comparison}).

Secondly, our injection recovery procedure was more comprehensive, involving $\sim 72$ million injections to characterize our sensitivity across different subsets of our sample to probe the impact of spectral subtype on our detections. Crucially, this step did not stop at detection, but instead included the full vetting process used for real candidates (excluding the \texttt{OddEvenFlag}; refer to \autoref{sec:injection}), to closely mimic the true selection pipeline. This approach allows us to rigorously assess the performance of our detection and vetting steps and generate a realistic characterization of completeness across our sample space.

Thirdly, we perform detailed spectroscopic vetting of each surviving candidate using high-resolution observations from HPF, WINERED, and NEID. This is a key improvement over previous occurrence rate studies, which typically relied on statistical false positive probabilities (FPPs) calculate exclusively with TESS photometry. Our spectroscopic follow-up was critical to our final candidate list: of the 13 new GEMS candidates identified prior, none remained viable after follow-up observations. Comparing our occurrence for early M-dwarfs with those from \cite{gan_occurrence_2023}, and those for each of our spectral subtypes with those from \cite{bryant_occurrence_2023} in \autoref{fig:occrates_lit}, we can see the importance of this approach to obtain an accurate estimate of giant planet occurrence, especially in the regime of faint host stars that are susceptible to background contamination and poor false positive rate estimates. We note here that close-in giant planet candidates identified in Kepler photometry also suffered from a high FP rate, although the FP rate measured in this work is considerably higher \citep{Santerne2012}.

Finally, we adopt a different approach to estimate the occurrence rate and subsequent errors, which we will improve upon in our full 200 pc search. For example, \citet{bryant_occurrence_2023} calculated $n_{pl}$ per bin as

\begin{align}
    n_{pl} = \sum_{i=1}^{N_{cands}} 1 - \text{FPP}_i
\end{align}

where FPP is the false positive probability of a given signal. 

\citet{bryant_occurrence_2023} calculated the FPP for each of their candidates using the \texttt{TRICERATOPS} package for a given candidate \citep{giacalone_vetting_2021}. \texttt{TRICERATOPS} is a statistical validation tool that uses transit photometry and stellar neighborhood information to determine the likelihood that a transit signal is caused by a false positive scenario (e.g., an eclipsing binary or background object) rather than a planet. We considered using \texttt{TRICERATOPS} to validate our own planet candidates; however, when testing the tool on a subset of confirmed GEMS within 200 pc using only our TGLC data, we found that it statistically validated 10 out of the 20 objects in this sample with a generous planet probability threshold of 50\%. This limitation stems from the relatively large point spread function of TESS (approximately 1 arcminute) and the correspondingly large photometric uncertainties for faint M-dwarfs, which reduce the ability of \texttt{TRICERATOPS} to exclude false positives. 

\citet{gan_occurrence_2023} employed a two-step approach to estimate the FPP for each of their candidates. First, they used \texttt{Forecaster} \citep{chen_probabilistic_2017}, which applies a probabilistic mass–radius relation to infer the likelihood that a transit signal arises from a low-mass star rather than a planet. Second, they estimated the probability of a candidate being a brown dwarf (BD) by dividing the number of known BDs by the number of known GEMS at the time, which does not account for detection biases, survey completeness, and the underlying occurrence rates of each population. 

For this work, we relied entirely on spectroscopic followup to determine concretely whether a candidate was an FP or a real planet. We consider this the gold standard for calculating robust occurrence rates.

\subsection{Future Work}
This search within a 100 pc M-dwarf sample is a pilot study to prepare for the analysis of a 200 pc M-dwarf sample, which represents a further eight-fold increase in sample size. Work beyond the 200 pc sample will aim to produce occurrence rates for giant exoplanets orbiting all M-dwarfs observed by TESS. The larger sample will also aid in adopting more sophisticated occurrence rate estimation techniques such as an HBM approach with astrophysical stellar dependent priors or the ABC method. 

This search also yielded 267 non-noise false positives at the manual vetting stage, including 142 circularized EBs, 19 eccentric EBs, 69 contact binaries, and two other transiting systems with periods beyond the scope of this study. Further work on these FPs will utilize spectroscopic observations, which coupled with TESS photometry, and ground-based photometric follow-up will allow us to confirm the false positive nature of these objects. These observations of double-lined EBs, which make up the majority of the FP sample, provide some of the best and most direct measurements of masses, radii, and limb darkening coefficients, which are especially useful in a regime where asteroseismology is not effective \citep{henry_character_2024}. These model-independent measurements of the M-dwarfs can then be used to compare and constrain theoretical models of low mass stars near the hydrogen-burning limit, a regime where models still show significant discrepancies from observations \citep{Baroch2018}, and will help to enable a more accurate characterization of the planets orbiting them \citep{benedict_solar_2016}. The sample of FPs will also allow us to help form statistics of close-in binaries around M-dwarf primaries, helping to inform the statistics of close companions to M-dwarf primaries, which will help reconcile the differences seen between planet occurrence rates derived from radial velocity surveys with transit surveys \citep{Moe2021}.

In addition to direct outgrowths of this study, this work emphasizes the need for future work across several different areas. \autoref{fig:occ_per} demonstrates the need for more precise constraints on GEMS with orbital periods longer than 10 days. Such constraints might be enabled by the upcoming \underline{PLA}netary \underline{T}ransits and \underline{O}scillations of stars mission (PLATO; \citet{Rauer2025}), which includes M-dwarfs in its target list ($M_* < 0.5 ~\mathrm{M_\odot}$) and a nominal mission life of $>4$ years. These estimates would help to determine whether an ice line occurrence rate turnover similar to that observed by \citet{fernandes_hints_2019} in FGK-star giant planets exists for GEMS as well. Furthermore, \autoref{fig:occ_rad} notes the lack of occurrence rate data for super-Neptunes orbiting M-dwarfs.

\section{Summary} \label{sec:summary}
As part of the \textit{Searching for GEMS} survey, we performed a pilot study of a distance-limited ($<$ 100 pc) sample of M-dwarfs observed by TESS using our new \texttt{TESS-miner} package. Each of our candidates was subjected to further exhaustive vetting, pixel-checks and transit fitting, following which we were left with 23 candidates. Of these, 7 were existing planets, 3 had previously been identified as astrophysical FPs by past research, and 13 were identified as astrophysical false positives based on spectroscopic observations. Based on this, we calculated occurrence rates for GEMS around early-type ($0.118\% \pm 0.068\%$), mid-type ($0.153\% \pm 0.069\%$), and late-type M-dwarfs ($0.036\% \pm 0.024\%$). We also calculated an overall GEMS occurrence rate across the complete M-dwarf sample to be $0.068\%\pm0.024\%$. Our estimates on occurrence of giant planets across the M-dwarf spectral type, when combined with literature estimates for FGK stars, enabled us to fit an exponential relation between giant planet occurrence rates and stellar mass, which confirms the rarity of GEMS due to the lower average mass protoplanetary disks on average around M-dwarfs. 

\begin{acknowledgments}

Resources supporting this work were provided by the (i) NASA High-End Computing Program through the NASA Center for Climate Simulation (NCCS) at Goddard Space Flight Center (ii) Pennsylvania State University's Institute for Computational and Data Sciences' (ICDS) Roar supercomputer, and (iii) the Resnick High Performance Computing Center, a facility supported by Resnick Sustainability Institute at the California Institute of Technology. This content is solely the responsibility of the authors and does not necessarily represent the views of the NCCS, ICDS, or Caltech.

These results are based on observations obtained with the Habitable-zone Planet Finder Spectrograph on the HET. We acknowledge support from NSF grants AST-1006676, AST-1126413, AST-1310885, AST-1310875,  ATI 2009889, ATI-2009982, AST-2108512, AST-2108801 and the NASA Astrobiology Institute (NNA09DA76A) in the pursuit of precision radial velocities in the NIR. The HPF team also acknowledges support from the Heising-Simons Foundation via grant 2017-0494. 

The Hobby-Eberly Telescope is a joint project of the University of Texas at Austin, the Pennsylvania State University, Ludwig-Maximilians-Universität München, and Georg-August Universität Gottingen. The HET is named in honor of its principal benefactors, William P. Hobby and Robert E. Eberly. The HET collaboration acknowledges the support and resources from the Texas Advanced Computing Center. We thank the Resident astronomers and Telescope Operators at the HET for the skillful execution of our observations with HPF. We would like to acknowledge that the HET is built on Indigenous land. Moreover, we would like to acknowledge and pay our respects to the Carrizo \& Comecrudo, Coahuiltecan, Caddo, Tonkawa, Comanche, Lipan Apache, Alabama-Coushatta, Kickapoo, Tigua Pueblo, and all the American Indian and Indigenous Peoples and communities who have been or have become a part of these lands and territories in Texas, here on Turtle Island.

WIYN is a joint facility of the University of Wisconsin-Madison, Indiana University, NSF's NOIRLab, the Pennsylvania State University, Purdue University, University of California-Irvine, and the University of Missouri. 

This research was carried out, in part, at the Jet Propulsion Laboratory, California Institute of Technology, under a contract with the National Aeronautics and Space Administration (80NM0018D0004).

The authors are honored to be permitted to conduct astronomical research on Iolkam Du'ag (Kitt Peak), a mountain with particular significance to the Tohono O'odham. Data presented herein were obtained at the WIYN Observatory from telescope time allocated to NN-EXPLORE through the scientific partnership of NASA, the NSF, and NOIRLab (program IDs 2023B-438370 and 2024A-103024, both with PI Kanodia).

This paper uses WINERED data gathered with
the 6.5 meter Magellan Telescope located at Las Campanas Observatory, Chile. We are grateful to Shogo Otsubo, Yuki Saragaku, and Tomomi Takeuchi (Kyoto-Sangyo University), and to Noriyuki Matsunaga (University of Tokyo) of the WINERED team, as well as the staff of Las Campanas Observatory, for their support during the WINERED observations. WINERED was developed by the University of Tokyo and the Laboratory of Infrared High-resolution Spectroscopy, Kyoto Sangyo University, under the financial support of KAKENHI (Nos. 16684001, 20340042, and 21840052) and the MEXT Supported Program for the Strategic Research Foundation at Private Universities (Nos. S0801061 and S1411028). The observing run in 2023 June was partly supported by KAKENHI (grant No. 18H01248) and JSPS Bilateral Program Number JPJSBP120239909. Andrew McWilliam thanks and acknowledges receipt of a Carnegie Venture Grant, kindly provided by the Carnegie Institution for Science, in order to purchase equipment required to adapt, install and support
WINERED on the Magellan Clay telescope.

This work has made use of data from the European Space Agency (ESA) mission Gaia (\url{https://www.cosmos.esa.int/gaia}), processed by the Gaia Data Processing and Analysis Consortium (DPAC, \url{https://www.cosmos.esa.int/web/gaia/dpac/consortium}). Funding for the DPAC has been provided by national institutions, in particular the institutions participating in the Gaia Multilateral Agreement.

Some of the data presented in this paper were obtained from MAST at STScI. Support for MAST for non-HST data is provided by the NASA Office of Space Science via grant NNX09AF08G and by other grants and contracts.

This work includes data collected by the TESS mission, which are publicly available from MAST. Funding for the TESS mission is provided by the NASA Science Mission directorate. 

This research made use of the (i) NASA Exoplanet Archive, which is operated by Caltech, under contract with NASA under the Exoplanet Exploration Program, (ii) SIMBAD database, operated at CDS, Strasbourg, France, (iii) NASA's Astrophysics Data System Bibliographic Services, and (iv) data from 2MASS, a joint project of the University of Massachusetts and IPAC at Caltech, funded by NASA and the NSF.

This research has made use of the SIMBAD database, operated at CDS, Strasbourg, France, 
and NASA's Astrophysics Data System Bibliographic Services.

This research has made use of the Exoplanet Follow-up Observation Program website, which is operated by the California Institute of Technology, under contract with the National Aeronautics and Space Administration under the Exoplanet Exploration Program.

CIC acknowledges support by an appointment to the NASA Postdoctoral Program at the Goddard Space Flight Center, administered by ORAU through a contract with NASA.
We thank Thomas Barclay for sharing the latest Roman GBTDS pointings. 
We thank Corey Beard for assistance in developing and testing sub-modules related to differences between odd and even transits.

Finally, we thank the anonymous referee for their careful review and thoughtful suggestions, which have significantly improved the clarity and quality of this work.
\end{acknowledgments}

\facilities{TESS, HET (HPF), WIYN (NEID), Magellan:Clay (WINERED)}

\software{
\texttt{pickle} \citep{van1995python},
\texttt{astropy} \citep{astropy_2013,astropy_collaboration_astropy_2018,astropy_2022},
\texttt{barycorpy} \citep{kanodia_python_2018},
\texttt{HxRGproc} \citep{ninan_habitable-zone_2018}, 
\texttt{ipython} \citep{perez_ipython_2007},
\texttt{K2fov} \citep{2016ascl.soft01009M},
\texttt{matplotlib} \citep{hunter_matplotlib:_2007},
\texttt{MRExo} \citep{kanodia_mass-radius_2019, kanodia_beyond_2023},
\texttt{numpy} \citep{harris_array_2020},
\texttt{pandas} \citep{mckinney-proc-scipy-2010},
\texttt{scipy} \citep{oliphant_python_2007, scipy},
\texttt{batman} \citep{kreidberg_batman_2015},
\texttt{lightkurve} \citep{lightkurve},
\texttt{wotan} \citep{Wotan},
\texttt{WARP} \citep{2024PASP..136a4504H} , 
\texttt{tglc} \citep{han_tess-gaia_2023},
\texttt{TRICERATOPS} \citep{giacalone_vetting_2021},
\texttt{LEO-vetter} \citep{Kunimoto_2024},
\texttt{juliet} \citep{Espinoza2019},
\texttt{dynesty} \citep{speagle_dynesty_2020}
}

\bibliography{manual_bib, MyLibrary}{}
\bibliographystyle{aasjournal}

\appendix \label{app}

\begin{figure*}[hbp]
    \centering
    \includegraphics[width=\textwidth]{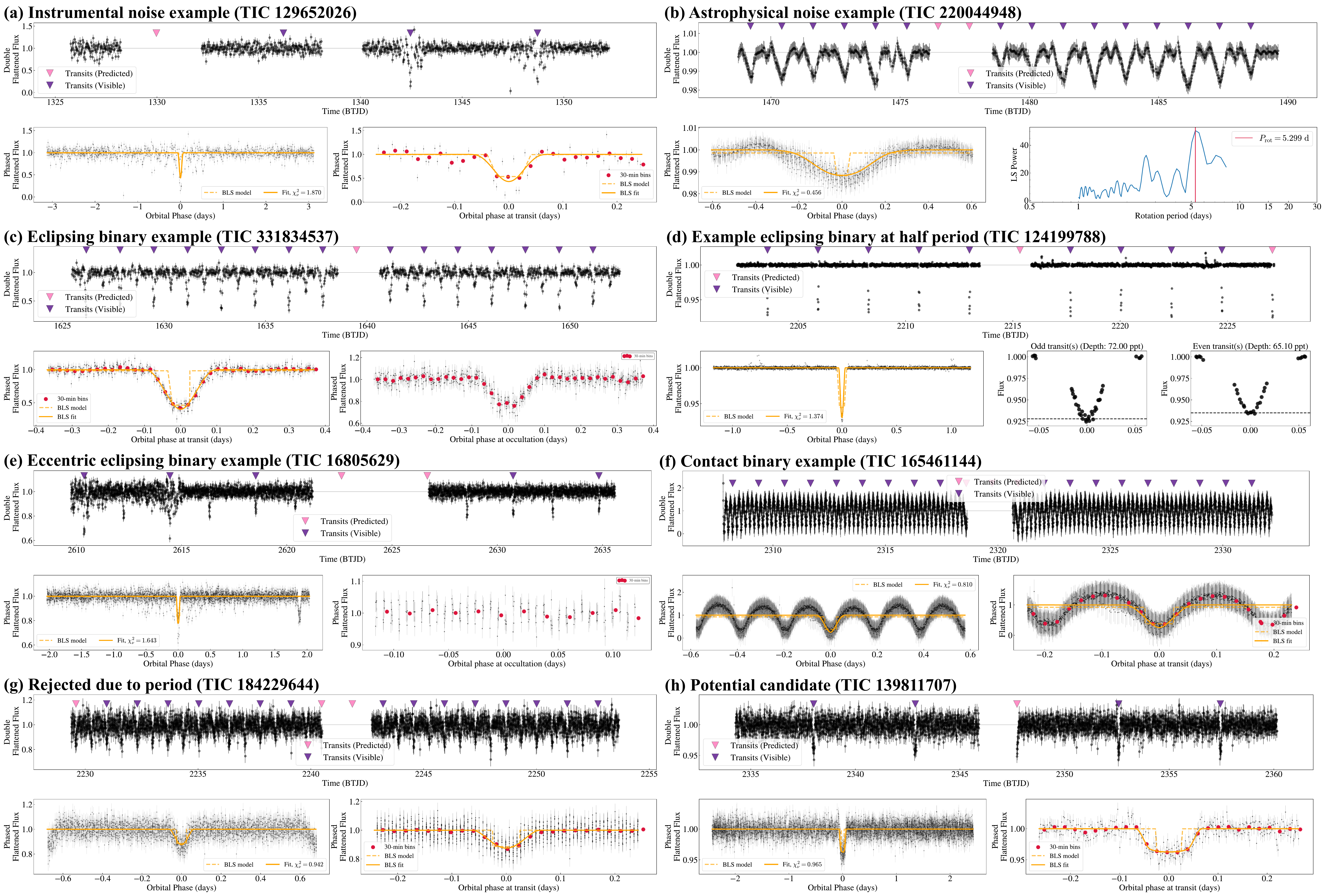}
    \caption{Example detrended and phase-folded light curves vetted manually in Section \ref{sec:vetting}. All figures are direct outputs of the \texttt{TESS-miner} package. \textbf{a.} shows an example of instrumental noise being falsely caught by the BLS search as a planet candidate. \textbf{b.} shows an example of the stellar rotation signal doing the same. \textbf{c.} shows an example of an EB FP, where the BLS search retrieved twice the period, therefore invalidating the \texttt{OddEvenFlag}. \textbf{d.} shows an example of an EB FP, where the BLS search retrieved half the period. \textbf{e.} shows an example of an eccentric EB FP, where the phase-folded light curve reveals an occultation offset from the expectation for a circular orbit. \textbf{f.} shows an example of a contact binary EB FP. \textbf{g.} shows an example of an otherwise acceptable planet candidate where the true period is outside out scope of interest (either $P<1$ day or $P>10$ days). Finally, \textbf{h.} is an example of an acceptable candidate.}
    \label{fig:man_ex}
\end{figure*}

\autoref{fig:man_ex} shows an example of each of the 7 types of FPs identified by eye in Stage 2 of our analysis, along with an example of a light curve which survived Stage 2. 

\begin{figure*}[hbp]
    \centering
    \begin{subfigure}{0.7\textwidth}
        \centering
        \includegraphics[width=\textwidth]{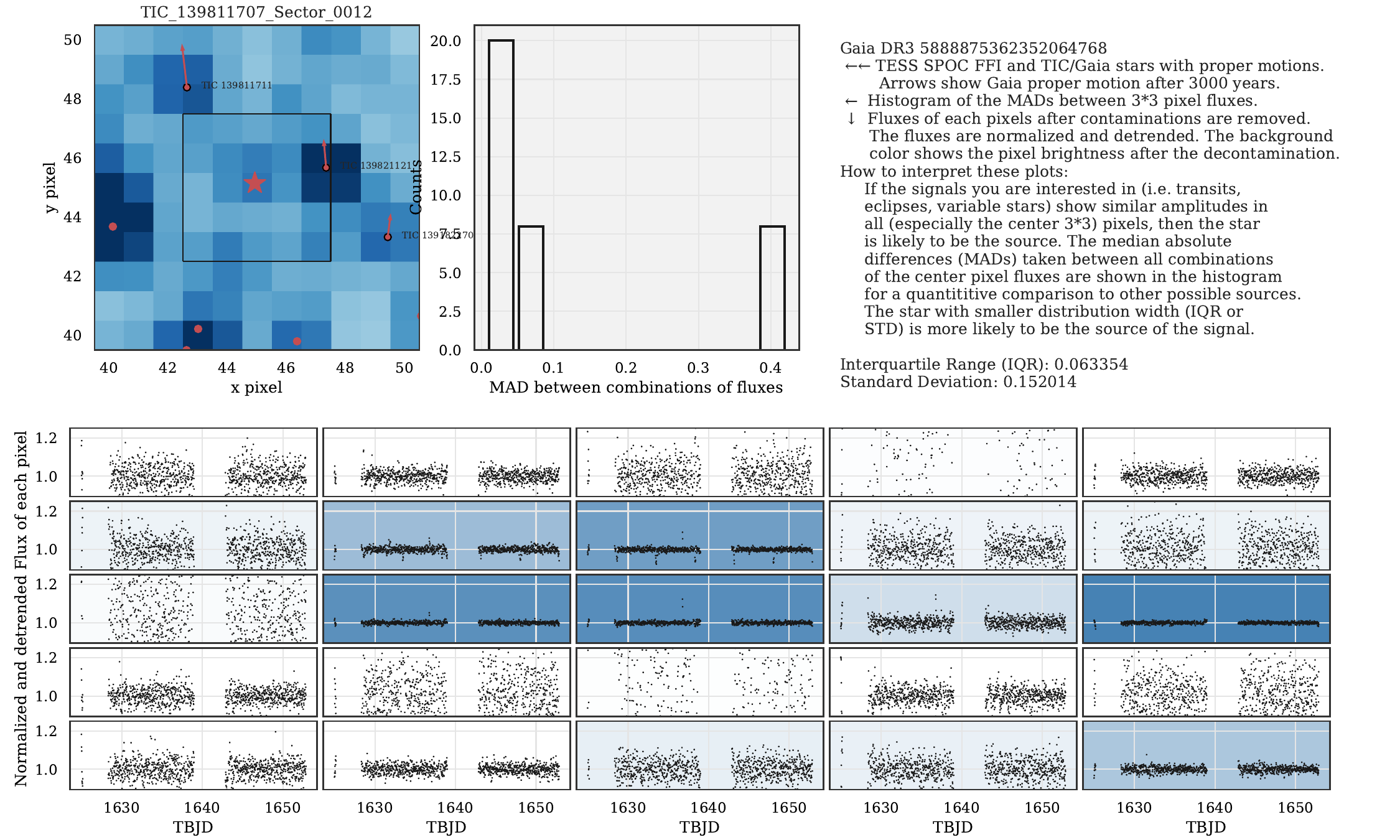}
        \caption{The pixel-pixel normalised light curves for TIC 139811797, Sector 12, one of our new planet candidates. The transit signal is visible in the pixels centred on the position of the target and do not appear to originate from any other star. For this reason, this object passed our pixel-pixel check.}
        \label{fig:pxpx1}
    \end{subfigure}
    
    \vspace{1cm} 

    \begin{subfigure}{0.7\textwidth}
        \centering
        \includegraphics[width=\textwidth]{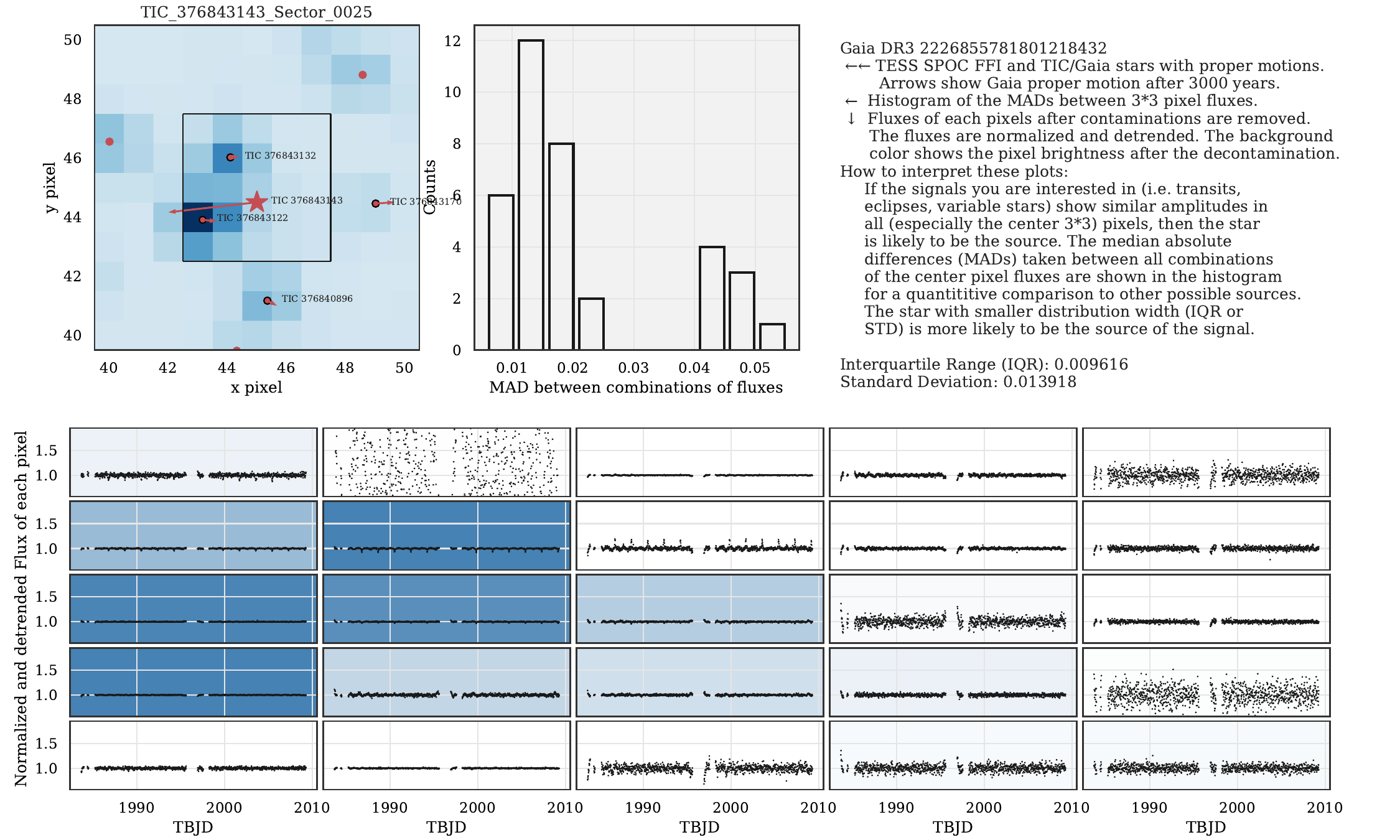}
        \caption{The pixel-pixel normalised light curves for TIC 376843143, Sector 25. The transit signal is visible in the pixels centred on TIC 376843132 and not the target star. For this reason, this objected did not pass our pixel-pixel check.}
        \label{fig:pxpx2}
    \end{subfigure}
    
    \caption{\textbf{Top:} an example of a pixel-pixel check pass, and \textbf{Bottom:} an example of a pixel-pixel check failure. These figures are generated by the \texttt{tglc} package.}
    \label{fig:pxpx}
\end{figure*}

\newpage
\autoref{fig:pxpx} shows an example of a TGLC pixel grid where light from the target star is uncontaminated by other sources (a) and an example of a pixel grid where this is not the case (b). In the latter case, a candidate is removed from consideration. This pixel-pixel check was performed during Stage 2 of analysis.

\newpage
\autoref{tab:tois} compiles each TOI within our input sample not explicitly noted in the main text of this work where $1.0~\mathrm{days} < P_{\text{TOI}} < 10.0$ days and $0.7 < R_{\text{TOI}}/R_J < 1.5$. We explain briefly in which stage each object was eliminated and why.

\begin{deluxetable}{c|c|p{10cm}}[H]
\tabletypesize{\normalsize}
\tablewidth{\linewidth}
\tablecaption{All TOI within our input sample and search criteria with which we have have not otherwise explicitly dealt in this work. \label{tab:tois}}
\tablehead{
\colhead{TIC} & 
\colhead{TOI} & 
\colhead{Details}
}
\startdata
146846569 & 734.01 & This object survived until Stage 2a. We removed it as a likely EB due to its V-shaped transit and evidence of a secondary eclipse, particularly in Sector 36. \\
67646988 & 1779.01 & This object also survived until the manual vetting stage, when we removed it due to its fitted $R_p/R_* = 0.35$ and V-shaped transit. \\
151728428 & 5850.01 & This object was classified as a Poor Fit during Stage 0. Several factors contributed to this classification. The object was only observed in one sector within the TESS PM and EM1 (Sector 54). Within that sector, an excess of noise led to a significant fraction of the light curve being cut from the analysis. Only one transit occurred within the remaining data, and consequently the \texttt{AntiTransitBLSFlag1} and the \texttt{RadiusFlag} were both triggered due to poor fitting. In addition, this object was retired as an SB1 by the TOI working group in May, 2024. \\
\enddata
\end{deluxetable}

\end{document}